\documentclass[12pt]{article}
\usepackage{amsmath,bbm,amsfonts}
\usepackage{graphicx,psfrag,epsf}
\usepackage{enumerate}
\usepackage{amsthm,amssymb}
\usepackage{caption}
\usepackage{natbib}
\usepackage{subcaption}
\usepackage{mathtools}
\usepackage{titlesec}
\usepackage{authblk}
\usepackage[latin1]{inputenc}
\usepackage{mathrsfs}
\usepackage{url} 
\usepackage{algpseudocode} 
\usepackage{enumitem}

\def\dif{{\rm d}}
\def\re{{\rm e}}
\def\ri{{\rm i}}
\def\tr{{\rm tr}}
\def\tk{\tilde \kappa}
\def\ta{\tilde a}
\def\tb{\tilde b}
\def\sT{\scriptscriptstyle{(T)}}

\def\equaldist{\stackrel{\rm d}{=}}

\usepackage{xparse}
\NewDocumentCommand{\ceil}{s O{} m}{%
  \IfBooleanTF{#1} 
    {\left\lceil#3\right\rceil} 
    {#2\lceil#3#2\rceil} 
}

\NewDocumentCommand{\floor}{s O{} m}{%
  \IfBooleanTF{#1} 
    {\left\lfloor#3\right\rfloor} 
    {#2\lfloor#3#2\rfloor} 
}

\newtheorem{theorem}{Theorem}

\newtheorem{corollary}{Corollary}
\newtheorem{proposition}{Proposition}
\newtheorem{assumption}{Assumption}

\begin{document}

\def\theequation{S\arabic{equation}}
\def\spacingset#1{\renewcommand{\baselinestretch}%
{#1}\small\normalsize} \spacingset{1}


\title{Wavelet Spectra for Multivariate Point Processes}
\date{}
\author[1]{Edward A. K. Cohen}
\author[2]{Alexander J. Gibberd} 
\affil[1]{{\small Department of Mathematics, Imperial College London, South Kensington Campus, London~SW7~2AZ,~U.K. }} 
\affil[2]{{\small Department of Mathematics and Statistics, Lancaster University, Bailrigg, Lancaster~LA1~4YF,~U.K.}} 

\maketitle

\spacingset{1.45} 

\setlength{\parindent}{0cm}
\setlength{\parskip}{0.3cm}

\begin{abstract}
	Wavelets provide the flexibility to analyse stochastic processes at different scales. Here, we apply them to multivariate point processes as a means of detecting and analysing unknown non-stationarity, both within and across data streams. To provide statistical tractability, a temporally smoothed wavelet periodogram is developed and shown to be equivalent to a multi-wavelet periodogram. Under a stationary assumption, the distribution of the temporally smoothed wavelet periodogram is demonstrated to be asymptotically Wishart, with the centrality matrix and degrees of freedom readily computable from the multi-wavelet formulation. Distributional results extend to wavelet coherence; a time-scale measure of inter-process correlation. This statistical framework is used to construct a test for stationarity in multivariate point-processes. The methodology is applied to neural spike train data, where it is shown to detect and characterise time-varying dependency patterns.
\end{abstract}

\section{Introduction}
We adopt the construction of \cite{Hawkes1971} which presents a $p$-dimensional multivariate point process ($p\geq1$) as a counting
vector $N(t)\equiv \{N_{1}(t),\ldots,N_{p}(t)\}^{\rm T}$ where the random
element $N_{i}(t)$ $(i=1,...,p)$ states the number of events of type $i$ over
the interval $(0,t]$. Its first order properties are characterized by its rate $\lambda(t)\in\mathbb{R}^p$, defined as $\lambda(t)\equiv E\{\dif N(t)\}/\dif t$ where $\dif N(t) = N(t+\dif t)-N(t)$, and its second order properties at times $s$ and $t$ characterized by its covariance density matrix
\begin{equation*}
\Gamma(s,t)=E\{\dif N(s)\dif N^{\rm T}(t)\}\big/(\dif t\ \dif s)-\lambda(s)\lambda^{\rm T}(t)\;.
\end{equation*}
Process $N(t)$ is second-order stationary (henceforth referred to simply as ``stationary'') if $\lambda(t)$ is constant for all $t$ and $\Gamma(t,s)$ depends only on $\tau = s-t$. In this setting we will denote the covariance density matrix $\Gamma(\tau)$. 

The spectral domain provides a rich environment for representing this second order structure and is based on the fact that stationary stochastic processes can be considered a composite of subprocesses operating at different frequencies. The spectral density matrix of a stationary point process is the Fourier transform of its covariance density matrix \citep{Bartlett1963a}, namely
\begin{equation*}
S(f)=\mathrm{diag}(\lambda)+\int_{-\infty}^{\infty}\Gamma(\tau)\re^{-i 2\pi f \tau}\dif \tau,\quad -\infty<f<\infty.\label{eq:spectral_density}
\end{equation*}
A fundamental summary of the second order relationship between a pair of component processes, $N_i(t)$ and $N_j(t)$ say, is their coherence defined as
\begin{equation}
\label{coh}
\rho^2_{ij}(f) = \frac{|S_{ij}(f)|^2}{S_{ii}(f)S_{jj}(f)}.
\end{equation}
This provides a normalized measure on $[0,1]$ of the correlation structure between the processes in the frequency domain. For time series data, it has been used extensively in several disciplines, including climatology, oceanography and medicine. For event data, it has been an important tool in neuroscience for the analysis of neuron spike train data.

Estimation of the coherence can be achieved by substituting smoothed spectral estimators into (\ref{coh}). Failure to smooth (i.e. simply using the periodogram) will result in a coherence estimate of one for all frequencies, irrespective of whether correlation exists between the pair of processes or not. Tractability of the coherence estimator's distribution is crucial for principled statistical testing and dependent on the smoothing procedure used \citep{Walden2000}.

Often, stochastic processes do not conform to the assumptions of stationarity. This might occur through simple first-order trends in the underlying data generating process, or more typically, complex changes in the second (or higher) order structure of the process. This renders classical Fourier methods obsolete and demands more flexible non-parametric methodology, with wavelets forming a natural basis with which to analyse non-stationary behaviour at different scales. 

For a wavelet $\psi(t)$, the continuous wavelet transform at scale $a>0$ and translation (or time) $b\in\mathbb{R}$ of $N(t)$, observed on the interval $(0,T]$, is defined by \cite{Brillinger1996} as
\begin{equation}
w(a,b) = a^{-1/2}\int_0^{T} \psi^{*}\{(t-b)/a\}\dif N(t)\;,
\label{eq:wavelet_cwt}
\end{equation}
where $^*$ denotes the complex conjugate.
The $i$th element of this stochastic integral is computed as 
$w_i(a,b) = \sum_{k=1}^{N_i(T)}\psi^{*}_{a,b}(s_{i,k})$, 
where $s_{i,1},...,s_{i,N_i(T)}$ are the ordered event times of $N_i(t)$ and $\psi_{a,b}(t)\equiv a^{-1/2}\psi\{(t-b)/a\}$. Thus, working with the continuous time process is possible if the finite set of event times are known. The wavelet periodogram is subsequently defined as
%
$W(a,b) = w(a,b)w^{\rm H}(a,b),\;$
where $^{\rm H}$ denotes the complex conjugate transpose. 
%

As is the case with the Fourier periodogram, smoothing is required for two reasons. Firstly to control variance, and secondly to give meaningful values of the wavelet coherence estimator. Wavelet coherence is an analogue of coherence which provides a normalized measure on $[0,1]$ of the correlation between a pair of processes in time-scale space. It is defined as 
\begin{equation*}
\gamma^2_{ij}(a,b) = \frac{|\Omega_{ij}(a,b)|^2}{\Omega_{ii}(a,b)\Omega_{jj}(a,b)},
\end{equation*}
where $\Omega$ is a smoothed version of $W$. In the time series setting, wavelet coherence has been extensively applied in a wide range of disciplines \citep[e.g.][]{Torrence99,Grinsted04}. Understanding the distributional properties of these smoothed coherence estimators is vital for rigorous statistical analysis and testing. In the Gaussian discrete-time setting the asymptotic distribution of coherence is widely studied \citep{Cohen2010a,Cohen2010}, however, the point-process case has received little attention. 

There are a wide range of ways in which non-stationarity can occur. Hence, rather than assume a specific model of non-stationarity, we here propose to study the properties of the temporally smoothed wavelet periodogram and coherence for stationary point-processes, thus providing a framework in which we deploy methods for exploratory data analysis and formal tests for detecting non-stationary. 

\section{Temporally smoothed wavelet periodogram}
\subsection{Formulation}
\begin{assumption}
	\label{assump:wav}
	Wavelet $\psi(t)$ is a real or complex valued continuous function that satisfies (i) $\int_{-\infty}^{\infty}\psi(t)\dif t =0$, (ii) $\|\psi\| = 1$, and (iii) the admissibility condition $\int_{-\infty}^\infty f^{-1}|\Psi(f)|^2 \dif f<\infty$, where $\Psi$ is the Fourier transform of $\psi(t)$.
\end{assumption}

\begin{assumption}
	\label{assump:h}
	Smoothing function $h(t)$ is a non-negative, symmetric function supported and continuous on $(-1/2,1/2)$, and normalized such that $\int_{-\infty}^{\infty}h(t)\dif t=1$.
\end{assumption}
Let $\psi(t)$ and $h(t)$ satisfy Assumptions \ref{assump:wav} and \ref{assump:h}, respectively. We define the \emph{temporally smoothed wavelet periodogram} as
\begin{eqnarray}
\Omega\left(a,b\right) & = & \int_{-\infty}^\infty h_{\xi}(u-b)W(a,u)\dif u,\label{eq:TSWP}
\end{eqnarray}
where $h_\xi\equiv \xi^{-1}h(t/\xi)$ with $\xi>0$ controlling the level of smoothing. This is a wavelet analogue to Welch's \emph{weighted overlapping sample averaging} spectral estimator for stationary time series \citep{Welch1967,Carter1987}. 
It will prove convenient for the level of smoothing to scale with $a$, and we therefore let $\xi = \kappa a$, with $\kappa>0$. 

For a particular choice of $\kappa$, and defining the Hermitian kernel function (at scale $a=1$) as

\begin{equation}
K(s,t) = \int_{-\infty}^{\infty}h_{\kappa}(u)\psi(s-u)\psi^*(t-u)\dif u\;,\label{eq:TSWP_kernel}
\end{equation}
the temporally smoothed wavelet periodogram in (\ref{eq:TSWP}) can be expressed as
\[
\Omega(a,b)\equiv\int_{0}^{T}\int_{0}^{T}K_{a,b}(s,t)\dif N(t)\dif N^{\rm T}(s)\;,
\]
where
$
K_{a,b}(s,t)=a^{-1}K\{(s-b)/a,(t-b)/a\}
$.
The $(i,j)$th element of $\Omega(a,b)$ is computed as
\begin{equation*}
\Omega_{ij}(a,b)=\sum_{k=1}^{N_i(T)}\sum_{k'=1}^{N_j(T)}K_{a,b}(s_{i,k},s_{j,k'})\;.
\end{equation*}

%
Given a choice for $h(t)$ and $\kappa$, the form of $K(s,t)$ will depend on $\psi(t)$. Throughout this paper, we use the examples of the complex valued Morlet wavelet and the real valued Mexican hat wavelet. These are examples of wavelets for which $K(s,t)$ is analytically tractable. 

\subsection{Practical implementation}
\begin{figure}[t!]
	\begin{center}\includegraphics[width=.8\linewidth]{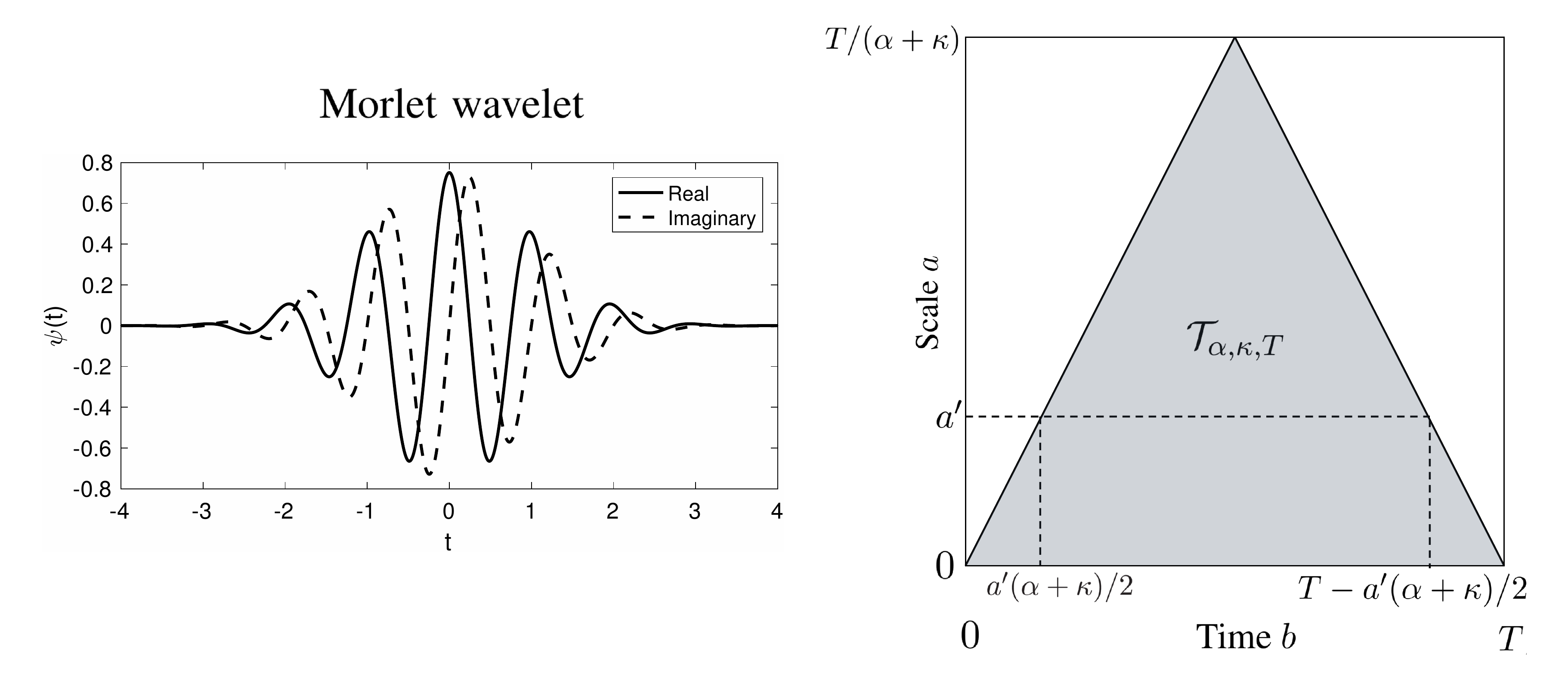}\caption{\label{triangle}The Morlet wavelet and the valid region for analysis $\mathcal{T}_{\alpha,\kappa,T}$. Note this has been plotted with time $b$ on the horizontal axis and scale $a$ on the vertical axis, as is convention}
	\end{center}
\end{figure}

For continuous time wavelet analysis, the wavelets themselves are
often non-compactly supported. However, the region of significant support is typically well localized and a close approximation
to $w(a,b)$ can be obtained through utilising the approximating
wavelet $$\bar\psi(t) = \left\{\begin{array}{cc}
\psi(t) & |t|< \alpha/2 \\ 
0 & \text{otherwise.}
\end{array} \right.$$
For example, the Morlet wavelet $\psi(t) = \pi^{-1/4}\re^{-t^2/2}\re^{\ri 2\pi t}$ shown in Fig. \ref{triangle} has infinite support but can be well approximated by $\bar\psi(t)$ for $\alpha=8$.
In practice, to speed up computation, it can make sense to use the
approximating wavelet as only a subset of the data is required to compute the wavelet transform. From herein, to simplify notation, will we use $\psi(t)$ to represent both the original and approximating wavelet, assuming that $\alpha$ is chosen appropriately.

In a finite data setting we are restricted to regions of the time-scale space in which we can fairly evaluate (\ref{eq:wavelet_cwt}) without the consequences of edge effects at either ends of the data. 
These issues are compounded when smoothing across time, for a smoothing window $h_\kappa(t)$ with ${\rm supp}(h) = (-\kappa/2,\kappa/2)$, the effective size of support for $K(s,t)$ is $\alpha+\kappa$, therefore we restrict ourselves to values of $a$ and $b$ for which ${\rm supp}(K_{a,b})= (b-a(\alpha+\kappa)/2,b+a(\alpha+\kappa)/2)\times(b-a(\alpha+\kappa)/2,b+a(\alpha+\kappa)/2)\subseteq (0,T]\times (0,T]$. This defines an isosceles triangle $\mathcal{T}_{\alpha,\kappa,T}\subset\mathbb{R}^2$ with vertices $(0,0)$, $(0,T)$ and $(a_{\rm max}(T),T/2)$, where $a_{\rm max}(T) = T/(\alpha+\kappa)$. This is an adaptation to the \emph{cone of influence} \citep[p. 215]{mallat2008} that also mitigates for smoothing distances.
%
%
In practice, a positive minimum value of $a$ should be imposed to ensure a reasonable amount of event data exists in the smoothing range. 

\section{Multi-wavelet representation}
\label{mwrep}
\subsection{Formulation}
Given $K(s,t)$ is continuous and non-negative definite by construction, associated with kernel $K(s,t)$ is the Hermitian linear operator $T_K$ defined as 
$
[T_K f](s) = \int_{-\infty}^{\infty}K(s,t)f(t)\dif t.
$
It follows from Mercer's
Theorem \citep{Mercer1909} that 
$
K(s,t) = \sum_{l=0}^{\infty}\eta_{l} \varphi_{l}(s)\varphi^{*}_{l}(t)
$
where $\{\varphi_{l}(t);\  l=0,1,...\}$ are the orthonormal eigenfunctions of $T_K$ with non-zero eigenvalues $\{\eta_{l};\  l=0,1,...\}$ ordered in decreasing size. Noting that $\tr (T_K): = \int_{-\infty}^{\infty}K(t,t)\dif t = 1$, it follows that $\sum_{l=0}^{\infty} \eta_{l} = 1$. From here on, we refer to $\{\varphi_{l}(t);\  l=0,1,...\}$ as the eigenfunctions of $K(s,t)$.
The following proposition shows that these orthonormal eigenfunctions are themselves wavelets. 
\begin{proposition}
	\label{phiisawavelet}
	Let $\psi(t)$ satisfy Assumption \ref{assump:wav}, $h(t)$ satisfy Assumption \ref{assump:h}, and for $\kappa>0$ the corresponding non-negative definite kernel $K(s,t)$ have eigenfunctions $\{\varphi_{l}(t);\  l=0,1,...\}$. Every eigenfunction $\varphi_{l}(t)$ with a non-zero eigenvalue is a wavelet that satisfies the conditions of Assumption \ref{assump:wav}.
\end{proposition}
We adopt the term \emph{eigen-wavelets} for the functions $\{\varphi_l(t); \ l=0,1,...\}$. 

Turning our attention back to the temporally smoothed wavelet periodogram, it is straightforward to show
\begin{equation*}
\int_{-\infty}^\infty K_{a,b}(s,t)\varphi_l\{(t-b)/a\}\dif t = \eta_l\varphi_l\{(s-b)/a\}. \label{eq:kernel_integral}
\end{equation*}
Thus, the scaled and shifted versions $\varphi_{l,a,b}(t)=a^{-1/2}\varphi_{l}\{(t-b)/a\}$, $l=0,1,\ldots$ of the eigen-wavelets are themselves the eigenfunctions of $K_{a,b}$, and again from Mercer's theorem
$
K_{a,b}(s,t) = \sum_{l=0}^{\infty}\eta_{l} \varphi_{l,a,b}(s)\varphi^{*}_{l,a,b}(t).
$
The temporally smoothed wavelet periodogram can thus be represented as 
\begin{equation}
\label{tswpsum}
\Omega(a,b) = \sum_{l=0}^\infty \eta_{l}v_l(a,b)v_l^{\rm H}(a,b),
\end{equation}
where $v_l(a,b) = \int_0^T \varphi_{l,a,b}(t)\dif N(t)$ is the continuous wavelet transform of $N(t)$ at scale $a$ and translation $b$ with respect to eigen-wavelet $\varphi_l(t)$. Therefore the temporally smoothed wavelet periodogram is equivalent to the weighted sum of wavelet spectra arising from the orthonormal eigen-wavelet system. This is analogous to multitapering \citep{Thomson1982} and comparisons can also be drawn with the multi-wavelet spectrum of \cite{Cohen2010}. In that setting, multiple orthogonal wavelets were derived in \cite{Olhede2002} from a time-frequency concentration problem, whereas here we have shown they can be generated by any arbitrary wavelet $\psi(t)$ and smoothing window $h(t)$. 

The representation in (\ref{tswpsum}) will be crucial for deriving the distributional results in Section \ref{sec:statprops}, as well as offering computational speed-up. In particular, we will make use of the following proposition which shows the effective frequency response of the eigen-wavelet system is equal to the frequency response of the generating wavelet $\psi(t)$.
%
%
\begin{proposition}
	\label{eigenwavfreq}
	Let $\psi(t)$ satisfy Assumption \ref{assump:wav}, $h(t)$ satisfy Assumption \ref{assump:h}, and for $\kappa>0$ the corresponding non-negative definite kernel $K(s,t)$ have eigenfunctions $\{\varphi_{l}(t);\  l=0,1,...\}$ and eigenvalues $\{\eta_l;\ l=0,1,...\}$. It holds that $\sum_{l}\eta_{l}|\Phi_{l}(f)|^{2}=|\Psi(f)|^{2}$ where $\Phi_l$ and $\Psi(f)$ are the Fourier transforms of $\varphi_l(t)$ and $\psi(t)$, respectively.
\end{proposition}

In general, closed form expressions for the eigen-wavelets $\{\varphi_l(t);\ l=0,1,\ldots\}$ will be unobtainable and numerical procedures need to be used to find the solutions of 
$
\int_{-\infty}^\infty K(s,t)\varphi(t)\dif t = \eta\varphi(s).
$
Details for an implementation of the Nystrom method for doing just this can be found in Appendix 1.

\subsection{Worked example}

\begin{figure}
	\begin{center}
		\includegraphics[width=0.7\linewidth]{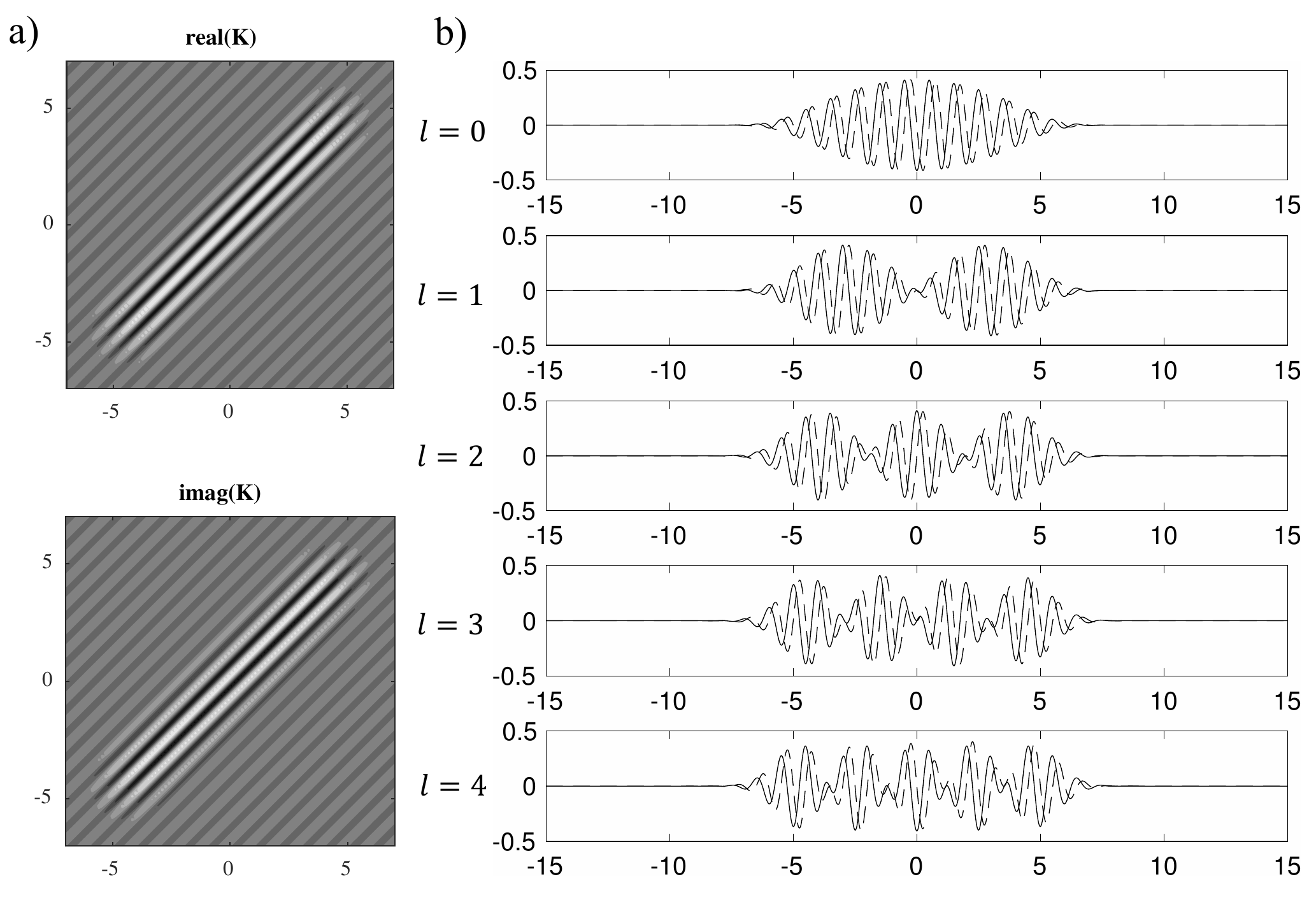}\\
		\includegraphics[width=0.7\linewidth]{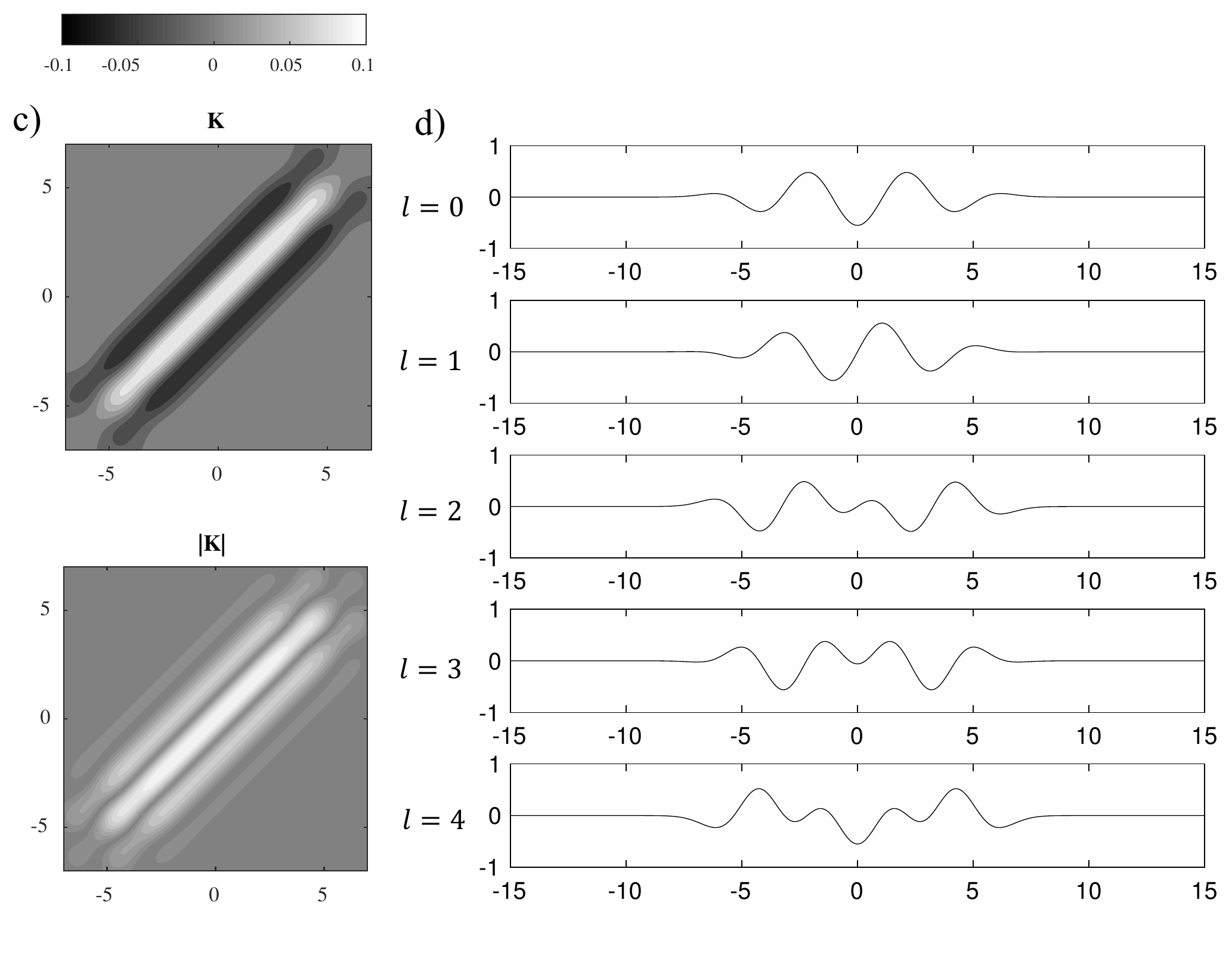}\caption{\label{eigenwavelets}The kernel and first five eigen-wavelets for the Morlet wavelet (panels a,b) and Mexican hat wavelet (panels c,d) subject to a rectangular smoothing window of width $\kappa=10$. In panel b) the solid and dashed line represent the real and imaginary components respectively.}
	\end{center}
\end{figure}

The Morlet wavelet 
%
can be seen as a complex sinusoid enveloped with a Gaussian window, and therefore the wavelet transform at scale $a>0$ and translation $b$ is the Fourier transform of the tapered process, localized at $b$ and evaluated at frequency $1/a$. The temporally smoothed wavelet periodogram using a rectangular smoothing function
\begin{equation}
\label{swindow}
h(t) = \left\{\begin{array}{lll}
1 &\qquad & -1/2<t<1/2 \\
0 & \qquad &\mbox{otherwise,}
\end{array}\right.
\end{equation}
emits kernel
$
K(s,t) = k(s,t)\re^{-\ri2\pi(t-s)},
$
where
\begin{equation*}
k(s,t) = (2\kappa)^{-1}\re^{-\left(t-s\right)^{2}}[\mathrm{erf}\{\kappa-(t+s)\}+\mathrm{erf}\{\kappa+(t+s)\}]
\end{equation*}
and $\mathrm{erf}(x) = \pi^{-1/2}\int_{-x}^{x}\exp(-t^2)\dif t$ is the Gauss error function. The real part of this kernel is shown in Fig. \ref{eigenwavelets}a. 

The function $k(s,t)$ is itself a real valued non-negative kernel with its own set of real valued orthonormal eigenfunctions $\{\phi_l(t);\ l=0,1,...\}$ and associated eigenvalues $\{\eta_l;\ l=0,1,...\}$. It follows that $\varphi_l(t) = \re^{\ri 2\pi t}\phi_l(t)$ is an eigenfunction of $K(s,t)$ with corresponding eigenvalue $\eta_l$ and hence $\{\varphi_l(t)=\re^{\ri 2\pi t}\phi_l(t);l=0,1,...\}$ is the eigen-wavelet system emitted by the Morlet wavelet with a rectangular smoothing function. The first five of these eigen-wavelets for $\kappa=10$ are shown in Fig. \ref{eigenwavelets}b. This eigen-wavelet system follows the same spirit of the generating Morlet wavelet, with themselves being complex sinusoids enveloped by a taper. Thus, performing a continuous wavelet transform with one of the eigen-wavelets is equivalent to a time localized tapered Fourier transform evaluated at frequency $1/a$, and the temporally smoothed wavelet periodogram as represented in (\ref{tswpsum}) is equivalent to a time localized multitaper spectral estimator. As means of a comparison, the kernel and associated eigen-wavelets of the Mexican hat wavelet using a rectangular smoothing function are shown in Fig. \ref{eigenwavelets}c and Fig. \ref{eigenwavelets}d, respectively. 


\section{Statistical Properties under Stationarity}
\label{sec:statprops}

\subsection{Preliminaries}

Let us define the $k$th order cumulant $q$ of the differential process as 
\begin{equation*}
q_{i_1,...,i_k}(t_1,...,t_{k})\dif t_1\cdots \dif t_k\equiv {\rm cum}\{\dif N_{i_1}(u_1),..., \dif N_{i_k}(u_k)\}\;.
\end{equation*}
The following mixing condition (Assumption 2.2 in \citet{Brillinger1972}) is sufficient for the asymptotic results that follow. It ensures that dependency structure in the point process decays at a sufficient rate for central limit arguments to be invoked. 
\begin{assumption}
	\label{assumption1}
	The $p$-dimensional point process $N(t)$ is strictly stationary, i.e. $q_{i_1,...,i_k}(t_1+t,...,t_{k}+t) = q_{i_1,...,i_k}(t_1,...,t_k)$, and we set $r_{i_1,...,i_k}(u_1,...,u_{k-1}) = q_{i_1,...,i_k}(u_1,...,u_{k-1},0)$. Furthermore, all moments exist, the cumulant function satisfies
	$$
	\int_{-\infty}^\infty \cdots \int_{-\infty}^\infty  |r_{i_1,...,i_k}(u_1,...,u_{k-1})|\dif u_1\cdots \dif u_{k-1}< \infty,
	$$
	for $i_1,...,i_k = 1,...,p$ and $k=2,3,...$, and
	$$
	\int_{-\infty}^{\infty}|u||r_{i_1,i_2}(u)|\dif u < \infty,
	$$
	for $i_1,i_2 = 1,...,p$.
\end{assumption}


The distributional results differ slightly depending on whether a real valued wavelet (e.g. Mexican hat) or complex valued wavelet (e.g. Morlet) is chosen. We present the results for a complex valued wavelet and relegate the derivation for a real valued wavelet to the Supplementary Material. 
\begin{assumption}
	\label{assumption2c}
	Wavelet $\psi(t)$ is complex valued, satisfies Assumption \ref{assump:wav} and has approximating support $(-\alpha/2,\alpha/2)$ for some finite $\alpha>0$. Furthermore, there exists a finite $C$ such that $\int|\psi(t+u)-\psi(t)|\dif t<C|u|$ for all real $u$, and it is orthogonal to its complex conjugate, i.e. $\int_{-\infty}^{\infty}\psi(t)\psi^*(t)\dif t = 0$.
\end{assumption}
The Morlet wavelet is an example of a complex valued wavelet that satisfies Assumption \ref{assumption2c}. 
\begin{assumption}
	\label{assumption3}
	Smoothing function $h(t)$ satisfies Assumption \ref{assump:h} and furthermore there exists a finite $C'$ such that $\int|h(t+u)-h(t)|\dif t<C'|u|$.
\end{assumption}
%
For wavelet $\psi(t)$ with Fourier transform $\Psi(f)$, its central frequency is defined as $f_{0}:=\int_{0}^{\infty}f|\Psi(f)|^{2}\dif f$ \citep{Cohen2010a}. The central frequency of $\psi_{a,b}$ is therefore $f_0/a$ and can be interpreted as the central analysing frequency of the wavelet at scale $a$. For example, the Morlet wavelet has a central frequency of $f_0 = 1$ and the Mexican hat wavelet has a central frequency of (approx.) $f_0 = 0.21$. It immediately follows from Proposition \ref{eigenwavfreq} that the central frequency of the eigen-wavelet system is $f_0$.

\subsection{Asymptotic distributional results}

We allow the wavelet to scale with $T$ by defining $\psi^{\sT}(t) = \{(\alpha+\kappa)/T\}^{-1/2}\psi\{t(\alpha+\kappa)/T\}$, and appropriately normalize the scale and translation parameters as $\ta = a(\alpha + \kappa)/T$ and $\tb = b/T$, respectively. Under this rescaling (\ref{eq:wavelet_cwt}) becomes
\begin{equation*}
w(a,b) = w^{\sT}(\tilde a,\tilde b) = \tilde{a}^{-1/2}\int_{0}^T \psi^{*\sT} \{(t-\tilde bT) / \tilde a \} \dif N(t),
\end{equation*}
and the normalized temporally smoothed wavelet periodogram is defined as
\begin{equation*}
\Omega^{\sT}(\ta,\tb) = \int_{-\infty}^{\infty}h^{\sT}_{\kappa \tilde a}(u)W^{\sT}(\ta,u)\dif u,
\end{equation*}
where $h^{\sT}(t) = T^{-1}h(t/T)$. For any $T$, the valid region of analysis is normalized to $\tilde{\mathcal{T}}_{\alpha,\kappa}$, an isosceles triangle with vertices $(0,0)$, $(0,1)$ and $(1,1/2)$ whose interior contains all valid pairs of $(\ta,\tb)$. Asymptotic results are presented for any fixed point $(\ta,\tb)\in \tilde{\mathcal{T}}_{\alpha,\kappa}$ as $T\rightarrow\infty$. In doing so, we define the frequency $f_{\ta} = f_0/(\ta T) = f_0/a$. 

\begin{proposition}
	\label{prop3}
	Let $N(t)$ be a $p$-dimensional stationary process with spectral density matrix $S(f)$. Let $\psi(t)$ be a wavelet satisfying Assumption \ref{assump:wav} and let $h(t)$ be a smoothing function satisfying Assumption \ref{assump:h}. For all $\kappa>0$ and for all $(\tilde a,\tilde b)\in \tilde{T}_{\alpha,\kappa}$,
	$$E\{\Omega^{\sT}(\tilde a,\tilde b) \}=E\{W^{\sT}(\tilde a,\tilde b)\} =  \int_{-\infty}^{\infty}\ta|\Psi^{\sT}(\ta f)|^2 S(f)\dif f$$
	and $E\{\Omega^{\sT}(\tilde a,\tilde b)\} = S(f_{\tilde a}) + O(T^{-2})$ as $T\rightarrow\infty$.
\end{proposition}

In the following theorem, $\mathcal{N}^{\mathcal{C}}_p(\mu,\Sigma)$ denotes the (circular) $p$-dimensional complex normal distribution with mean $\mu$ and covariance matrix $\Sigma$.
\begin{theorem}
	\label{thm:1c}
	Let $N(t)$ be a $p$-dimensional stationary process satisfying Assumption \ref{assumption1} with spectral density matrix $S(f)$, and let $\psi(t)$ be a wavelet with central frequency $f_0$ satisfying Assumption \ref{assumption2c}.  The continuous wavelet transform $w^{\sT}(\ta,\tb)$ is asymptotically $\mathcal{N}^{\mathcal{C}}_p\{0,S(f_{\ta})\}$ as $T\rightarrow\infty$, for all $(\ta,\tb)\in\tilde{\mathcal{T}}_{\alpha,\kappa}$.
\end{theorem}

Let $\mathcal{W}^{\mathcal{C}}_p(n,\Sigma)$ denote the $p$-dimensional complex Wishart distribution with $n$ degrees of freedom and centrality matrix $\Sigma$. 
\begin{theorem}
	\label{thm:2c}
	Let $N(t)$ be a $p$-dimensional stationary process satisfying Assumption \ref{assumption1} with spectral density matrix $S(f)$. Let $\psi(t)$ be a wavelet with central frequency $f_0$ satisfying Assumption \ref{assumption2c}, let $h(t)$ be a smoothing function satisfying Assumption \ref{assumption3}, and for $\kappa>0$ let $\{\eta_l;\ l=0,1,\}$ be the eigenvalues of the kernel $K(s,t)$ defined in (\ref{eq:TSWP_kernel}). The temporally smoothed wavelet periodogram $\Omega^{\sT}(\ta,\tb)$ is asymptotically $(1/n)\mathcal{W}^{\mathcal{C}}_p\{n,S(f_{\ta})\}$ as $T\rightarrow\infty$ for all $(\ta,\tb)\in \tilde{\mathcal{T}}_{\alpha,\kappa}$, where $n=1/\left(\sum_{l=1}^\infty\eta_l^2\right)$. 
\end{theorem}	
The following distributional result for the wavelet coherence is now immediate from Theorem \ref{thm:2c} and \cite{Goodman1963}. We let $_2 F_1(\alpha_1,\alpha_2;\beta_1;z)$ denote the hypergeometric function with 2 and 1 parameters $\alpha_1$, $\alpha_2$ and $\beta_1$ and scalar argument $z$.
\begin{corollary}
	\label{cor:1c}	
	Under the conditions of Theorem \ref{thm:2c}, the temporally smoothed wavelet coherence $
	\gamma_{ij}^2(\ta,\tb) = |\Omega^{\sT}_{ij}(\ta,\tb)|^2/\{\Omega^{\sT}_{ii}(\ta,\tb)\Omega^{\sT}_{jj}(\ta,\tb)\}$ between component processes $N_i(t)$ and $N_j(t)$ ($i\neq j$) asymptotically has density function $$g_{\gamma^2}(x) =(n-1)(1-\rho^2)^n(1-x)^{n-2}\; _2 F_1(n,n;1;\rho^2x),$$
	where $\rho^2$ is shorthand for $\rho_{ij}^2(f_{\ta})$, the spectral coherence between $N_i(t)$ and $N_j(t)$ at frequency $f_{\ta}$.
\end{corollary}
In the case of the rectangular smoothing function given in (\ref{swindow}), the effective degrees of freedom $n$ scale linearly with $\kappa$ according to the following proposition.
\begin{proposition}
	\label{prop4}
	Let $\psi(t)$ satisfying Assumption \ref{assumption2c}, let $h(t)$ be the rectangular smoothing function given in (\ref{swindow}), and for $\kappa>0$ let corresponding kernel $K(s,t)$ have ordered eigenvalues $\{\eta_{l};\  l=0,1,...\}$. Provided $\kappa>\alpha$, then
	$
	n = (\sum_{l=0}^{\infty}\eta_l^2)^{-1} = \kappa \{\int^{\infty}_{-\infty} |\mathcal{P}(x)|^2\dif x\}^{-1} ,
	$ 
	where $\mathcal{P}(x) \equiv \int_{-\infty}^\infty\psi(t)\psi^*(t-x)\dif t$.
\end{proposition}

\section{Test for stationarity}
Consider testing the null hypothesis $H_0$ that states $N(t)$ is a stationary process, against the alternative hypothesis $H_A$ that states $H_0$ is not true. Under $H_0$ and from Proposition \ref{prop3} it is true that $E\{\Omega(a,b)\}$ is constant in $b$. We therefore consider testing for stationarity at different scales.

Consider a smoothing parameter $\tk = \kappa T^c$ where $\kappa>0$ and $0<c<1/2$. From Proposition \ref{prop4} we have degrees of freedom $n$ in Theorem \ref{thm:2c} being $O(T^c)$. With a slight reworking of the normalized framework of Section \ref{sec:statprops}, we set $\psi^{\sT}(t) = \{(\alpha+\tk)/T\}^{-1/2}\psi\{t(\alpha+\tk)/T\}$ and appropriately normalize the scale and translation parameters as $\ta = a(\alpha + \tk)/T$ and $\tb = b/T$. This again normalizes the valid region of analysis to $\tilde{\mathcal{T}}_{\alpha,\kappa}$ for all $T$.

For convenience, we perform a dyadic partition of the time-scales space, performing a test at each scale in the set $\{\ta_j=2^{-j};j=1,...,J\}$.  At scale $\ta_j$, we partition time into $2^j$ non-overlapping equal size segments, each centred at time points $\{\tb_{j,k} = (2k-1)/(2^{j+1});k=1,...,2^j\}$ and each the width of the approximate support of the wavelet at that scale. 

\begin{proposition}
	Let $\psi(t)$ satisfy Assumption \ref{assumption2c} and $h(t)$ satisfy Assumption \ref{assumption3}. Then, for any $\kappa>0$ and $j>0$, $\Omega^{\sT}(\ta_j,\tb_{j,1}),...,\Omega^{\sT}(\ta_j,\tb_{j,2^j})$ are asymptotically independent.
\end{proposition}

Our test at scale $\tilde a_j$ therefore becomes a test of the null hypothesis
\begin{center}
	$H_j:$ $E\{\Omega^{\sT}(\ta_j,\tb_1)\} =  ... =   E\{\Omega^{\sT}(\ta_j,\tb_{2^j})\} = \Omega_j$,
\end{center}
where $\Omega_j$ is unspecified. We construct a likelihood ratio test based on the asymptotic distribution of $\Omega^{\sT}(\ta,\tb)$ stated in Theorem \ref{thm:1c}.


\begin{proposition}
	\label{lemma1}
	Let $B_1,...,B_K$ be independent samples where $B_i\sim(1/n)\mathcal{W}^{\mathcal{C}}_p(n,\Sigma_i)$ ($i=1,...,K$). The likelihood ratio test statistic for the null hypothesis $H:\Sigma_1=...=\Sigma_K=\Sigma$, with unspecified $\Sigma$, is 
	$$
	\tilde\Lambda = K^{pK n}\frac{\prod_{i=1}^{K}{\rm det}(B_i)^n}{{\rm det}\left(\sum_{i=1}^KB_i\right)^{Kn}}.
	$$
	Furthermore, when $H$ is true, $-2\log (	\tilde\Lambda)$ is asymptotically $\chi^2_{f}$ where $f  = (K-1)p^2$.
\end{proposition}
In the following proposition, we let $\tilde\Lambda_0(\Sigma)$ be a random variable that is equal in distribution to $	\tilde\Lambda$ under the null hypothesis. 
\begin{proposition}
	Let $\tk = \kappa T^{c}$ where $\kappa>0$ and $0<c<1/2$, and define the test statistic for $H_j$ as
	$$
	\Lambda_j = K^{pK n}\frac{\prod_{i=1}^{K}{\rm det}\{\Omega^{\sT}(\ta_j,\tb_i)\}^n}{{\rm det}\left\{\sum_{i=1}^K\Omega^{\sT}(\ta_j,\tb_i)\right\}^{Kn}},
	$$
	where $K = 2^j$ and $n$ is as given in Theorem \ref{thm:2c}. Under $H_j$, $\Lambda_j\equaldist 	\tilde\Lambda_0(\Sigma) + o(1)$, where $\Sigma = E\{\Omega(a,b)\}$.
\end{proposition}
\begin{theorem}
	Let $\tk = \kappa T^{c}$ where $\kappa>0$ and $0<c<1/2$. Under $H_j$, $-2\log (\Lambda_j)$ is asymptotically $\chi^2_{\nu_j}$ where $\nu_j  = (2^j-1)p^2$. Specifically, ${\rm pr}\{-2\log (\Lambda_j)\leq x\} = {\rm pr}(\chi^2_{\nu_j}\leq x) + O(T^{-\beta})$ where $\beta = \min\{c,1/2-c\}$.
\end{theorem}
Thus, the rate of convergence to $\chi^2_{\nu_j}$ is optimal when $c=1/4$. 

Let $\psi^{\sT}_{j,k}(t)$ denote the wavelet at the $j$th scale and $k$th translation ($j=1,...,J$; $k=1,...,K$). Provided $\int_{-\infty}^{\infty}\psi_{j,k}^{\sT}(t)\psi_{l,m}^{*\sT}(t)\dif t = 0$ for all $(j,k)\neq(l,m)$ (this is only an approximation for the Morlet wavelet), the likelihood ratio test statistics will be independent of each other. Combining them for $H_1,...,H_J$, namely $\Lambda\equiv \prod_{j=1}^J \Lambda_j$, forms a test statistic for $H_0$. It follows that $-2\log(\Lambda)$ is asymptotically $\chi^2_\nu$, where $\nu = \sum_{j=1}^{J}\nu_j = p^2(2^{J+1}-2-J)$. 

\section{Real data example}

To give an example of the methods in practice, we analyse signalling regions within the \emph{lateral geniculate nucleus} of a mouse. Specifically, we consider a set of neurons examined in \cite{Tang2015}, where the authors are primarily concerned with analysing firing properties in order to understand how visual signals are encoded and transferred throughout the brain. To demonstrate the ability of our smoothed coherence estimator to operate with a single trial we consider only a single firing sequence from the paper. In this case, the mouse is shown a visual stimulus in the form of an liquid crystal display screen showing a sinusoidal monochromatic drifting grating with spatial stimulus at a frequency of 0.04 cycles per second and temporal flicker of 1Hz. The firing pattern is 7 seconds in length and represents data for cells 108 and 117 (these cells were picked for the example as they demonstrate relatively high firing rates). We use the Morlet wavelet with temporal smoothing parameter $\kappa=10$ and approximating support $\alpha=4$. For completeness, this example was performed using exact kernel sampling, however, an approximate computation based on the Nystrom method based method (see Appendix 1) provides visually indistinguishable results and p-values.

\begin{figure}[t!]
	\begin{center}\includegraphics[width=1\linewidth]{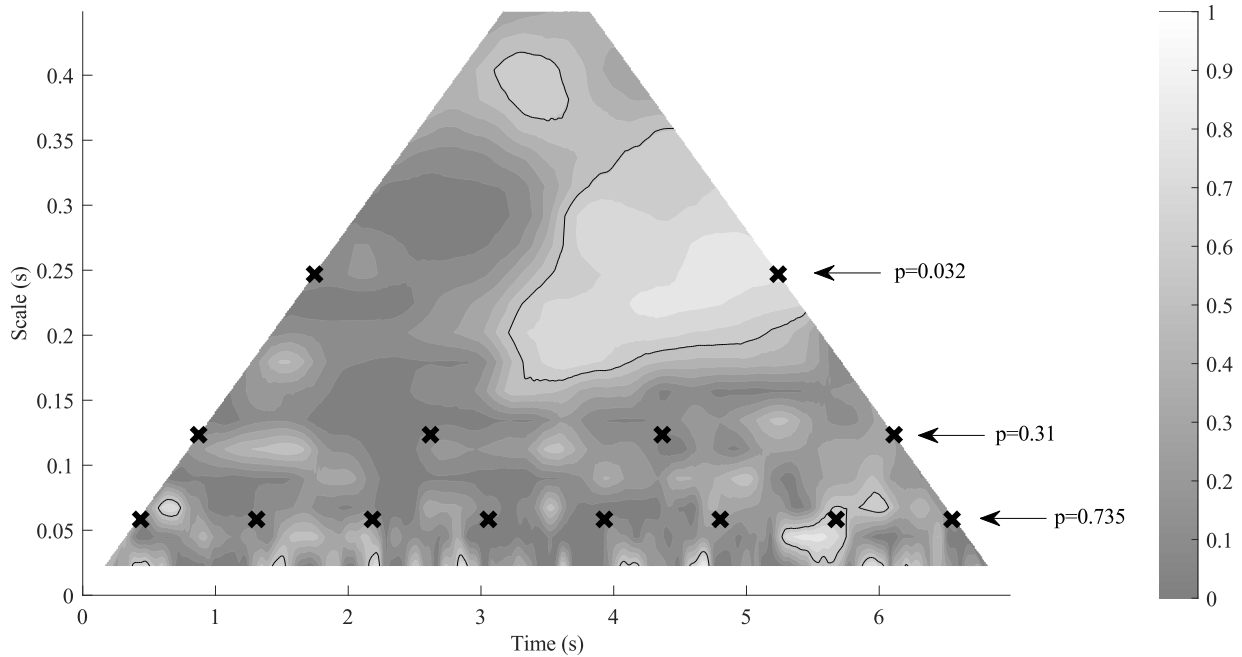}\caption{\label{mouse_data} Wavelet coherence estimated from the mouse neuron firing data with the Morlet wavelet. Black lines represent contours drawn at the 95th percentile of the distribution for zero coherence (see Corollary \ref{cor:1c}). Crosses indicate dyadic sampling points for the stationarity test, with arrows depicting the scale specific p-values. }
	\end{center}
\end{figure}

The analysis of the experimental data is provided in Fig. \ref{mouse_data}. Tests for stationarity were performed at scale levels $j=1,2,3$, with dyadic sampling points as marked by the crosses. With a p-value of 0.032, there is strong evidence that the process demonstrates non-stationary behaviour at the coarsest scale ($j=1$, corresponding to a scale of 0.25s and frequency of 4Hz through the relationship for a Morlet wavelet that $f=1/a$). However, there is little evidence to support non-stationarity existing at the finer scales ($j=2,3$). With the parameters specified, the 95th percentile of the distribution for zero coherence is 0.593 and is represented by the black contour line. This indicates that the non-stationarity at $j=1$ involves a change in the correlation between the two data streams half way through the experiment, with significant coherent signalling becoming present in the latter half. It is worth noting that whilst we can also see some peaks in the wavelet coherence at higher frequencies (scale 0.025s), the number of data points within the support of the kernel is limited. Thus, at this level, we should be careful to make inferences based on the asymptotic results.

\section*{Acknowledgement}
This work is funded by EPSRC grant EP/P011535/1. The authors would like to thank Leigh~Shlomovich, Department of Mathematics, Imperial College London for developing the Hawkes process simulation code, and Heather~Battey, Dean~Bodenham, and Andrew~Walden, Department of Mathematics, Imperial College London for stimulating conversations.

\section*{Supplementary material}
\label{SM}
Supplementary Material Section 1 contains the proofs to propositions and theorems presented here.  Supplementary Material Section 2 provides the results for real valued wavelets. Supplementary Material Section 3 provides verification of the results via simulation, as well as further supporting figures. It also contains a link to a MATLAB package for implementing the presented methods.

\appendix

\section*{Appendix 1}
\subsection*{Computing eigen-wavelets and eigenvalues}
\label{ap1}

The Nystrom method \citep[Chapter 1]{Kythe} is an efficient method for computing the eigenfunctions of kernel $K(s,t)$ for the multiwavelet representation described in Section \ref{mwrep}. We can approximate the integral using the quadrature rule to solve the approximate eigen-problem $\sum_{j=1}^n  w_j K(s,t_j)\tilde{\varphi}_l(t_j)=\tilde{\eta} \tilde{\varphi}_l(s)$ for a discrete set of values for $s$. The quadrature points $\{t_1,...,t_n\}$ ($n$ large) are regularly spaced across $(-(\alpha+\kappa)/2,(\alpha+\kappa)/2)$ and the weights are set to be $w_j=(\alpha+\kappa)/n$. For simplicity, the Nystrom points $\{ s_1,...,s_n\}$ are set to equal $\{t_1,...,t_n\}$. In matrix form, the eigen-problem now becomes
\[
KW\tilde\varphi=\tilde{\eta}\tilde\varphi\;,
\]
where $K$ is the $\mathbb{R}^{n\times n}$ matrix $(K(s_{i},t_{j}))$, $\tilde\varphi\equiv[\tilde{\varphi}(t_{1}),\ldots,\tilde{\varphi}(t_{n})]^{\rm T}$, and $W\equiv\mathrm{diag}(w_{1},\ldots,w_{n})$. Solving the above gives approximations to the first $n$ eigenvalues and eigen-wavelets of kernel $K(s,t)$.

Should it be required, the Nystrom extension of the sampled vector $\tilde\varphi_{l}=[\tilde{\varphi}(s_{1}),\ldots,\tilde{\varphi}(s_{n})]$
is the function 
\begin{equation*}
\tilde{\varphi}_{l}(x)=\tilde{\lambda}_{l}\sum_{j=1}^{n}w_{j}K(x,s_{j})\tilde{\varphi}_{l}(s_{j}).\label{eq:nystrom_extension}
\end{equation*}

The sum in (\ref{tswpsum}) is over an infinite set of (eigen-)wavelet periodograms. However, in practice, the size of the eigenvalues drop away rapidly indicating that the kernel can be accurately reconstructed using only a small number of its eigen-wavelets, hence (\ref{tswpsum}) can be approximated with only a small number of terms. For example, in the case of the $\kappa = 10$, the first nine eigenvalues contain 99.9\% (3.s.f.) of the overall energy.

\bibliographystyle{chicago}
\bibliography{bib}

\begin{thebibliography}{}

\bibitem[\protect\citeauthoryear{Bartlett}{Bartlett}{1963}]{Bartlett1963a}
Bartlett, M.~S. (1963).
\newblock The spectral analysis of point processes.
\newblock {\em Journal of the Royal Statistical Society. Series B\/}~{\em
  25\/}(2), 264--296.

\bibitem[\protect\citeauthoryear{Brillinger}{Brillinger}{1972}]{Brillinger1972}
Brillinger, D.~R. (1972).
\newblock {The spectral analysis of stationary interval functions}.
\newblock {\em Proceedings of the Sixth Berkeley Symposium on Mathematical
  Statistics and Probability, Volume 1: Theory of Statistics\/}, 483--513.

\bibitem[\protect\citeauthoryear{Brillinger}{Brillinger}{1996}]{Brillinger1996}
Brillinger, D.~R. (1996).
\newblock {Some uses of cumulants in wavelet analysis}.
\newblock {\em Journal of Nonparametric Statistics\/}~{\em 6}, 93--114.

\bibitem[\protect\citeauthoryear{Carter}{Carter}{1987}]{Carter1987}
Carter, G. (1987).
\newblock {Coherence and time delay estimation}.
\newblock {\em Proceedings of the IEEE\/}~{\em 75\/}(2), 236--255.

\bibitem[\protect\citeauthoryear{Cohen and Walden}{Cohen and
  Walden}{2010a}]{Cohen2010a}
Cohen, E. A.~K. and A.~T. Walden (2010a).
\newblock {A statistical analysis of Morse wavelet coherence}.
\newblock {\em IEEE Transactions on Signal Processing\/}~{\em 58\/}(3 PART 1),
  980--989.

\bibitem[\protect\citeauthoryear{Cohen and Walden}{Cohen and
  Walden}{2010b}]{Cohen2010}
Cohen, E. A.~K. and A.~T. Walden (2010b).
\newblock A statistical study of temporally smoothed wavelet coherence.
\newblock {\em IEEE Transactions on Signal Processing\/}~{\em 58\/}(6),
  2964--2973.

\bibitem[\protect\citeauthoryear{Goodman}{Goodman}{1963}]{Goodman1963}
Goodman, N. (1963).
\newblock Statistical analysis based on a certain multivariate complex
  {G}aussian distribution (an introduction).
\newblock {\em Annals of Mathematical Statistics\/}~{\em 34\/}(1), 152--177.

\bibitem[\protect\citeauthoryear{Grinsted, Moore, and Jevrejeva}{Grinsted
  et~al.}{2004}]{Grinsted04}
Grinsted, A, J., C.~Moore, and S.~Jevrejeva (2004).
\newblock Application of the cross wavelet transform and wavelet coherence to
  geophysical time series.
\newblock {\em Nonlinear Processes in Geophysics\/}~{\em 11\/}(5), 561--566.

\bibitem[\protect\citeauthoryear{Hawkes}{Hawkes}{1971}]{Hawkes1971}
Hawkes, A.~G. (1971).
\newblock Spectra of some self-exciting and mutually exciting point processes.
\newblock {\em Biometrika\/}~{\em 58\/}(1), 83--90.

\bibitem[\protect\citeauthoryear{Kythe and Puri}{Kythe and Puri}{2001}]{Kythe}
Kythe, P.~K. and P.~Puri (2001).
\newblock {\em Computational Methods for Linear Integral Equations}.
\newblock Springer.

\bibitem[\protect\citeauthoryear{Mallat and Peyr\'{e}}{Mallat and
  Peyr\'{e}}{2008}]{mallat2008}
Mallat, S. and G.~Peyr\'{e} (2008).
\newblock {\em {A Wavelet Tour of Signal Processing}\/} (3 ed.).
\newblock Elsevier Science and Technology.

\bibitem[\protect\citeauthoryear{Mercer}{Mercer}{1909}]{Mercer1909}
Mercer, J. (1909).
\newblock Functions of positive and negative type, and their connection with
  the theory of integral equations.
\newblock {\em Philosophical Transactions of the Royal Society A: Mathematical,
  Physical and Engineering Sciences\/}~{\em 209\/}(441-458), 415--446.

\bibitem[\protect\citeauthoryear{Olhede and Walden}{Olhede and
  Walden}{2002}]{Olhede2002}
Olhede, S.~C. and A.~T. Walden (2002).
\newblock Generalized {M}orse wavelets.
\newblock {\em IEEE Transactions on Signal Processing\/}~{\em 50\/}(11),
  2661--2670.

\bibitem[\protect\citeauthoryear{Tang, {Ardila Jimenez}, Chakraborty, and
  Schultz}{Tang et~al.}{2015}]{Tang2015}
Tang, S., S.~C. {Ardila Jimenez}, S.~Chakraborty, and S.~R. Schultz (2015).
\newblock {Visual receptive field properties of neurons in the mouse lateral
  geniculate nucleus}.
\newblock {\em Plos One\/}~{\em 11\/}(1), 1--34.

\bibitem[\protect\citeauthoryear{Thomson}{Thomson}{1982}]{Thomson1982}
Thomson, D.~J. (1982).
\newblock Spectrum estimation and harmonic analysis.
\newblock {\em Proceedings of the IEEE\/}~{\em 70\/}(9).

\bibitem[\protect\citeauthoryear{Torrence and Webster}{Torrence and
  Webster}{1999}]{Torrence99}
Torrence, C. and P.~Webster (1999).
\newblock Interdecadal changes in the {ESNO}-monsoon system.
\newblock {\em Journal of Climate\/}~{\em 12}, 2679--2690.

\bibitem[\protect\citeauthoryear{Walden}{Walden}{2000}]{Walden2000}
Walden, A.~T. (2000).
\newblock {A unified view of multitaper multivariate spectral estimation}.
\newblock {\em Biometrika\/}~{\em 87\/}(4), 767--788.

\bibitem[\protect\citeauthoryear{Welch}{Welch}{1967}]{Welch1967}
Welch, P. (1967).
\newblock The use of fast {F}ourier transform for the estimation of power
  spectra: a method based on time averaging over short, modified periodograms.
\newblock {\em IEEE Transactions on Audio Electroacoustics\/}~{\em 15}, 70--73.

\end{thebibliography}


\begin{thebibliography}{}

\bibitem[\protect\citeauthoryear{Barndorff-Nielsen and Cox}{Barndorff-Nielsen
  and Cox}{1989}]{OEBN1989}
Barndorff-Nielsen, O.~E. and D.~R. Cox (1989).
\newblock {\em Asymptotic Techniques for Use in Statistics}.
\newblock Chapman and Hall.

\bibitem[\protect\citeauthoryear{Brillinger}{Brillinger}{1972}]{Brillinger1972}
Brillinger, D.~R. (1972).
\newblock {The spectral analysis of stationary interval functions}.
\newblock {\em Proceedings of the Sixth Berkeley Symposium on Mathematical
  Statistics and Probability, Volume 1: Theory of Statistics\/}, 483--513.

\bibitem[\protect\citeauthoryear{Fisher}{Fisher}{1928}]{Fisher1928}
Fisher, R.~A. (1928).
\newblock The general sampling distribution of the multiple correlation
  coefficient.
\newblock {\em Proceedings of the Royal Society Series A\/}~{\em 121\/}(788).

\bibitem[\protect\citeauthoryear{Hawkes}{Hawkes}{1971}]{Hawkes1971}
Hawkes, A.~G. (1971).
\newblock Spectra of some self-exciting and mutually exciting point processes.
\newblock {\em Biometrika\/}~{\em 58\/}(1), 83--90.

\bibitem[\protect\citeauthoryear{Muirhead}{Muirhead}{1985}]{muirhead1985}
Muirhead, R.~J. (1985).
\newblock {\em {Aspects of Multivariate Statistical Theory}\/} (2 ed.).
\newblock John Wiley \& Sons, Inc.

\bibitem[\protect\citeauthoryear{Walden}{Walden}{2000}]{Walden2000}
Walden, A.~T. (2000).
\newblock {A unified view of multitaper multivariate spectral estimation}.
\newblock {\em Biometrika\/}~{\em 87\/}(4), 767--788.

\end{thebibliography}
\section*{Correspondence} Correspondence should be addressed to 

Edward Cohen \newline Department of Mathematics \newline Imperial College London \newline London SW7 2AZ. 

Email: e.cohen@imperial.ac.uk

\end{document}


\def\theequation{S\arabic{equation}}
\def\spacingset#1{\renewcommand{\baselinestretch}%
{#1}\small\normalsize} \spacingset{1}


\title{Supplementary Material \\
Wavelet Spectra for Multivariate Point Processes}
\date{}
\author{E.A.K. Cohen and A.J. Gibberd}
\maketitle

\spacingset{1.45} 

\setlength{\parindent}{0cm}
\setlength{\parskip}{0.3cm}

\section{Proofs}
Where relevant, $a$ and $b$ are such that $(a,b)\in\mathcal{T}_{\alpha,T}$ or $(a,b)\in\mathcal{T}_{\alpha,\kappa, T}$. This allows all integrals over $(0,T)$ to be replaced by integrals over the entire real line. 

We adopt throughout the convention that the Fourier transform $G(f)$ of a function $g(t)$ is defined as $G(f) = \int g(t)\re^{-\ri 2\pi f t}\dif t$, and the inverse Fourier transform defined as $g(t) = \int G(f)\re^{\ri 2\pi f t}\dif f$. To suppress notation, $\int$ implicitly denotes $\int_{-\infty}^{\infty}$.

While in the main manuscript it makes sense to present Proposition 1 before Proposition 2, for the purposes of proving these results, it makes sense to prove Proposition 2 first.

\subsection*{Proofs of Propositions 1-3}
\begingroup
\renewcommand\theproposition{2}
\begin{proposition}
	\label{eigenwavfreq}
	Let $\psi(t)$ satisfy Assumption 1, $h(t)$ satisfy Assumption 2, and for $\kappa>0$ the corresponding non-negative definite kernel $K(s,t)$ have eigenfunctions $\{\varphi_{l}(t);\  l=0,1,...\}$ and eigenvalues $\{\eta_l;\ l=0,1,...\}$. It holds that $\sum_{l}\eta_{l}|\Phi_{l}(f)|^{2}=|\Psi(f)|^{2}$ where $\Phi_l$ and $\Psi(f)$ are the Fourier transforms of $\varphi_l(t)$ and $\psi(t)$, respectively.
\end{proposition}
\begin{proof}
We define 
$$
\mathcal{K}(f,f') = \int\int K(s,t)\re^{-\ri 2\pi f s}\re^{-\ri 2\pi f't}\dif s \dif t.
$$
Using the representation
$$
K(s,t) = \int h_\kappa(u)\psi(s-u)\psi^*(t-u)\dif u
$$
gives 
$
\mathcal{K}(f,-f) = |\Psi(f)|^2,
$
recalling that $\int_{-\infty}^{\infty}h_\kappa(u)\dif u = 1$. Using the representation 
$$
K(s,t) = \sum_{l=1}^{\infty}\eta_l \varphi_l(s)\varphi_l^*(t)
$$
gives 
$
\mathcal{K}(f,-f) = \sum_{l=1}^{\infty}\eta_l |\Phi_l(f)|^2.
$ The required result follows.
\end{proof}
We can now easily proceed with the proof of Proposition 1.
\renewcommand\theproposition{1}
\begin{proposition}
	\label{phiisawavelet}
	Let $\psi(t)$ satisfy Assumption 1, $h(t)$ satisfy Assumption 2, and for $\kappa>0$ the corresponding non-negative definite kernel $K(s,t)$ have eigenfunctions $\{\varphi_{l}(t);\  l=0,1,...\}$. Every eigenfunction $\varphi_{l}(t)$ with a non-zero eigenvalue is a wavelet that satisfies the conditions of Assumption 1.
\end{proposition}
\begin{proof}
Each $\varphi_{l}(t)$ with non-zero eigenvalue will be real (complex) and continuous for real (complex) wavelet $\psi(t)$, by construction. Property (i): from the definition of $K(s,t)$ in (4), it immediately follows that $\int\int K(s,t) \dif s \dif t = 0$. Furthermore, with $K(s,t) = \sum_{l=0}^\infty \eta_l\varphi_l(s)\varphi^*_l(t)$, it follows that
$$
\int\int K(s,t) \dif s \dif t = \sum_{l=0}^\infty \eta_l \left|\int \varphi_l (t) \dif t\right|^2.
$$
Therefore $\left|\int \varphi_l (t) \dif t\right| = 0$ for all $\varphi_l(t)$ with positive eigenvalues.

Property (ii) is immediate from the construction of the eigenfunctions.

Property (iii) is immediate from Proposition 1 and the fact that $\psi(t)$ itself obeys the admissibility condition.
\end{proof}
\endgroup
\begin{proposition}
	\label{prop3}
	Let $N(t)$ be a $p$-dimensional stationary process with spectral density matrix $S(f)$. Let $\psi(t)$ be a wavelet satisfying Assumption 1 and let $h(t)$ be a smoothing function satisfying Assumption 2. For all $\kappa>0$ and for all $(\tilde a,\tilde b)\in \tilde{T}_{\alpha,\kappa}$,
$$E\{\Omega^{\sT}(\tilde a,\tilde b) \}=E\{W^{\sT}(\tilde a,\tilde b)\} =  \int_{-\infty}^{\infty}\ta|\Psi^{\sT}(\ta f)|^2 S(f)\dif f$$
and $E\{\Omega^{\sT}(\tilde a,\tilde b)\} = S(f_{\tilde a}) + O(T^{-2})$ as $T\rightarrow\infty$.
\end{proposition}
\begin{proof}
We can write 
$$
\Omega^{\sT}(\ta,\tb)=\int_{0}^{T}\int_{0}^{T}K^{\sT}_{\ta,\tb}(s,t)\dif N(t)\dif N^{\rm T}(s)
$$
where
$$
K^{\sT}(s,t) = \int h^{\sT}_{\kappa \tilde{a}}(u)\psi^{\sT}(s-u)\psi^{*\sT}(t-u)\dif u
$$
and $K^{\sT}_{\ta,\tb}(s,t) = \ta^{-1}K\{(s-\tb T)/\ta,(t-\tb T)/\ta\}$. If $\{\varphi_l(t);l=0,1,...\}$ are the eigenfunctions of $K(s,t)$, then $\{\varphi_l^{\sT}(t)=T^{-1/2}\varphi_l(t/T);l=0,1,...\}$ are the eigenfunctions of $K^{\sT}(s,t)$.
It follows that $E\{\Omega^{\sT}(\ta,\tb)\} = \sum_{l=0}^\infty \eta_l E\{v^{\sT}_l(\ta,\tb)v_l^{\sT \rm H}(\ta,\tb)\}$, where $v^{\sT}_l(\ta,\tb) =\ta^{-1/2} \int \varphi^{\sT}_l\{(t-\tb T)/\ta\}\dif N(t)$. Furthermore,
\begin{eqnarray*}E\{v^{\sT}_l(\ta,\tb)v_l^{\sT \rm H}(\ta,\tb)\} & =& \int\int \varphi^{\sT}_l\{(t-\tb T)/\ta\}\varphi^{*\sT}_l\{(s-\tb T)/\ta\}\Gamma(t,s)\dif t\dif s \\ & =& \int\int \varphi^{\sT}_l\{(t-\tb T)/\ta\}\varphi^{*\sT}_l\{(t-\tau-\tb T)/\ta\}\Gamma(\tau)\dif t\dif \tau \\ & =& \ta\int |\Phi^{\sT}_l(\ta f)|^2 S(f)\dif f,\end{eqnarray*} by using $\varphi^{\sT}(t) = \int \Phi^{\sT}(f)\re^{\ri 2\pi f t}\dif f$. Therefore, from Proposition~2
$$
E\{\Omega^{\sT}(\ta,\tb)\}=E\{W^{\sT}(\ta,\tb)\}= \ta\int |\Psi^{\sT}(\ta f)|^2S(f)\dif f.
$$
Taking the Taylor expansion of $S(f)$ around $f_{\ta}$ gives
\begin{eqnarray*}
\ta\int |\Psi^{\sT}(\ta f)|^2S(f)\dif f & = &\ta\int |\Psi^{\sT}(\ta f)|^2\{S(f_{\ta})+ (f-f_{\ta}) S'(f_{\ta}) + \frac{(f-f_{\ta})^2}{2!} S''(f_{\ta}) + ...\}\dif f \\
& = & S(f_{\ta}) + S''(f_{\ta})\ta\int |\Psi^{\sT}(\ta f)|^2\frac{(f-f_{\ta})^2}{2!}  \dif f + ... \\
& = & S(f_{\ta}) + O(T^{-2})
\end{eqnarray*}
from the Fourier transform uncertainty principle.
\end{proof}

\subsection*{Lemmas for Theorems 1-2}
The following Lemma is presented as Corollary 3.1 in \cite{Brillinger1972}. 
\begin{lemma}
	\label{lem:1}
	Let $N(t)$ be a $p$-dimensional point process satisfying Assumption 3, and let $\xi_1(t), ... , \xi_k(t)$ be continuous functions with finite support, then
	\begin{multline}\label{eqap}
	{\rm cum}\left\{\int\xi_1 (t_1) \dif N_{i_1}(t_1),...,\int\xi_k (t_k)\dif N_{i_k}(t_k)\right\} 
	= \sum_{l=1}^k \sum_{\alpha_1,...,\alpha_l = 1}^p \left(\prod_{j\in v_1}\delta_{\alpha_1i_j}\right)\cdots \left(\prod_{j\in v_l}\delta_{\alpha_li_j}\right)\\ \times\int\cdots\int \left\{\prod_{j\in v_1}\xi_j(\tau_1)\right\}\cdots \left\{\prod_{j\in v_l}\xi_j(\tau_l)\right\} q_{\alpha_1,...,\alpha_l}(\tau_1,...,\tau_l)\dif \tau_1,..,\dif \tau_l,
	\end{multline}
	where $\delta_{ij} = 1$ if $i=j$ and is zero otherwise. 
\end{lemma}
The first summation in (\ref{eqap}) does not have just $k$ terms, but instead extends over all partitions of $\{1,...,k\}$ of the form $(v_1,...,v_l)$. For example, for $k=3$, the $l=1$ partition is $(\{1,2,3\})$, the $l=2$ partitions are $(\{1\},\{2,3\})$, $(\{2\},\{1,3\})$ and $(\{3\},\{1,2\})$, and the $l=3$ partition is $(\{1\},\{2\},\{3\})$, resulting in 5 terms. To proceed, we will also require the following lemma.
\begin{lemma}
	\label{lem:2}
	Let $N(t)$ be a $p$-dimensional point process satisfying Assumption 3, and let $\psi_1(t)$ and $\psi_2(t)$ be a pair of orthogonal wavelets, each satisfying Assumption 4. The cumulant
	$$
	{\rm cum}\left\{\int\psi^{\sT}_1 (t_1) \dif N_{i_1}(t_1),\int\psi_2^{*\sT} (t_2)\dif N_{i_2}(t_2)\right\}$$ is $O(T^{-1})$.
\end{lemma}
\begin{proof}
	From Lemma \ref{lem:1}, the cumulant equals 
	$$ 
	\int\int \psi^{\sT}_{1}(t+u)\psi^{*\sT}_2(t)r_{i_1,i_2}(u)\dif u \dif t.
	$$
	The stated orthogonality of $\psi_1(t)$ and $\psi_2(t)$ implies $\int \psi_1^{\sT}(t)\psi_2^{*\sT}(t)\dif t = 0$ and Assumption 4 gives
	$\int\left|\psi_1^{\sT}(t+u)-\psi_1^{\sT}(t)\right|\dif t \leq T^{-1/2}C|u|$.
	Therefore,
	\begin{multline*}
	\left|	\int\int \psi^{\sT}_{1}(t+u)\psi^{*\sT}_2(t)r_{i_1,i_2}(u)\dif u \dif t\right| \\
	= 	\left|\int\int \psi^{\sT}_{1}(t+u)\psi^{*\sT}_2(t)r_{i_1,i_2}(u)\dif u \dif t - 	\int\int \psi^{\sT}_{1}(t)\psi^{*\sT}_2(t)r_{i_1,i_2}(u)\dif u \dif t.\right| \\
	\leq 	\int\int \left|\psi^{\sT}_{1}(t+u)\psi^{*\sT}_2(t)- \psi^{\sT}_{1}(t)\psi^{*\sT}_2(t)\right||r_{i_1,i_2}(u)|\dif u \dif t\\
	\leq 	\int\int \left|\psi^{\sT}_2(t)\right|\left|\psi^{\sT}_{1}(t+u)- \psi^{\sT}_{1}(t)\right||r_{i_1,i_2}(u)|\dif u \dif t \\ \leq T^{-1/2}A\int\int \left|\psi^{\sT}_{1}(t+u)- \psi^{\sT}_{1}(t)\right||r_{i_1,i_2}(u)|\dif u \dif t \leq
	T^{-1}AC\int|u||r_{i_1,i_2}(u)|\dif u,
	\end{multline*}
	where $A\equiv \max_t \left|\psi_2(t)\right|.$ Hence, by Assumption 3, the given cumulant is $O(T^{-1}).$
\end{proof}
\begin{lemma}
	\label{lem:3}
	Let $\psi(t)$ be a complex valued wavelet satisfying Assumption 4, and $h(t)$ a smoothing function satisfying Assumption 5. For $\kappa>0$, the eigen-wavelets of corresponding kernel $K(s,t)$ also satisfy Assumption 4.
\end{lemma}
\begin{proof}
The variability condition follows trivially from the variability condition on $\psi(t)$. The orthogonality condition also follows from the orthogonality condition on $\psi(t)$. Specifically, it is true that $\int\int K(s,t)K(s,t)\dif s \dif t =0$. Furthermore, $$\int\int K(s,t)K(s,t)\dif s \dif t=\sum_{l=0}^\infty\eta_l^2\left\{\int\varphi_l(t)\varphi_l(t)\dif t\right\}^2.$$ Therefore, $\int\varphi_l(t)\varphi_l(t)\dif t = 0$ for $l=0,1,...$, and $\varphi_{l}(t)$ is orthogonal to its complex conjugate.
\end{proof}
\subsection*{Proof of Theorem 1 and 2, and Proposition 4}
\begin{theorem}
	\label{thm:1c}
Let $N(t)$ be a $p$-dimensional stationary process satisfying Assumption 3 with spectral density matrix $S(f)$, and let $\psi(t)$ be a wavelet with central frequency $f_0$ satisfying Assumption 4.  The continuous wavelet transform $w^{\sT}(\ta,\tb)$ is asymptotically $\mathcal{N}^{\mathcal{C}}_p\{0,S(f_{\ta})\}$ as $T\rightarrow\infty$, for all $(\ta,\tb)\in\tilde{\mathcal{T}}_{\alpha,\kappa}$.
\end{theorem}
\begin{proof}
	We first verify the mean and covariance of $w^{\sT}(\ta,\tb)$ are as given. The mean is 
	\begin{align*}E\{w^{\sT}(\ta,\tb)\} & = E\left[\int_0^T \psi^{*\sT}\{(t-\tb T )/\ta\}\dif N(t)\right] \\ &= \int_0^T\psi^{*\sT}\{(t-\tb T)/\ta\}E\{\dif N(t)\} \\ & = \int_0^T\psi^{*\sT}\{(t-\tb T)/\ta\}\lambda(t)\dif t.\end{align*}
	Under the assumptions of the theorem, $N(t)$ is stationary, hence $\lambda(t)$ is constant for all $t$ and $E\{w(\ta,\tb)\} = 0$ as the wavelet integrates to zero. The asymptotic result for ${\rm cov}\{w^{\sT}(\ta,\tb)\} = E\{W^{\sT}(\ta,\tb)\}$ is given in Proposition 3. 
	
	Additionally, we are required to show ${\rm cov}\{w^{\sT}(\ta,\tb),w^{*\sT}(\ta,\tb)\}$ is asymptotically zero for a circular complex normal distribution. We note for a pair of complex random variables $W$ and $Z$, ${\rm cov}(W,Z^*) = {\rm cum}(W,Z)$. Therefore,
	\begin{multline*}
	{\rm cov}\{w_{i_1}^{\sT}(\ta,\tb),w^{*\sT}_{i_2}(\ta,\tb)\} \\ = {\rm cum}\left\{\ta^{-1/2}\int\psi^{*\sT} \left(\frac{t_1-\tb T}{\ta}\right) \dif N_{i_1}(t_1),\ta^{-1/2}\int\psi^{*\sT} \left(\frac{t_2-\tb T}{\ta}\right)\dif N_{i_2}(t_2)\right\}.
	\end{multline*}
Assumption 4 implies $\psi^{\sT}(t)$ is orthogonal to $\psi^{*\sT}(t)$. Setting $\psi_1(t) = \psi^{*\sT}(t)$ and $\psi_2(t)  = \psi^{\sT}(t)$ in Lemma \ref{lem:2} gives all the terms in ${\rm cov}\{w^{\sT}(\ta,\tb),w^{*\sT}(\ta,\tb)\}$ as $O(T^{-1})$. All first and second order cumulants are therefore asymptotically equal to those stated in the theorem.
	
	To conclude, we are required to show that all cumulants of order greater than two asymptotically go to zero. These cumulants can be written in the form $${\rm cum}\{w^{\sT}_{i_1}(\ta,\tb),...,w^{\sT}_{i_{k'}}(\ta,\tb),w^{*\sT}_{i_{k'+1}}(\ta,\tb),...,w^{*\sT}_{i_k}(\ta,\tb)\},$$ where $0 \leq k' \leq k$, $k>2$. In the quest to simplify notation, we present the $k'=0$ case, i.e. cumulants of the form ${\rm cum}\{w^{*\sT}_{i_{1}}(\ta,\tb),...,w^{*\sT}_{i_k}(\ta,\tb)\}$. The extension to cumulants that include both forms of conjugation is trivial.
		
	From Lemma \ref{lem:1},
	\begin{multline*}
{\rm cum}\{w^{*\sT}_{i_{1}}(\ta,\tb),...,w^{*\sT}_{i_k}(\ta,\tb)\}
	= \sum_{l=1}^k \sum_{\alpha_1,...,\alpha_l = 1}^p  \left(\prod_{j\in v_1}\delta_{\alpha_1i_j}\right)\cdots \left(\prod_{j\in v_l}\delta_{\alpha_li_j}\right)\\ \times\int\cdots\int \left[\ta^{-1/2}\psi^{\sT}\{(\tau_1-\tb T)/\ta\}\right]^{|v_1|}\cdots [\ta^{-1/2}\psi^{\sT}\{(\tau_l-\tb T)/\ta\}]^{|v_l|}q_{\alpha_1,...,\alpha_l}(\tau_1,...,\tau_l)\dif \tau_1,..,\dif \tau_l.
	\end{multline*}
	Through a change of variables and under Assumption 3 it follows that
	\begin{multline*}
{\rm cum}\{w^{*\sT}_{i_{1}}(\ta,\tb),...,w^{*\sT}_{i_k}(\ta,\tb)\}
	= \sum_{l=1}^k  \sum_{\alpha_1,...,\alpha_l = 1}^p  \left(\prod_{j\in v_1}\delta_{\alpha_1i_j}\right)\cdots \left(\prod_{j\in v_l}\delta_{\alpha_li_j}\right)\\ \int\cdots\int \left[\ta^{-1/2}\psi^{\sT}\{\left(t+u_1\right)/\ta\}\right]^{|v_1|}\cdots \left[\ta^{-1/2}\psi^{\sT}\left\{(t+u_{l-1})/\ta\right\}\right]^{|v_{l-1}|}\left[\ta^{-1/2}\psi^{\sT}\left(t\right/\ta)\right]^{|v_{l}|}\\ \cdot r_{\alpha_1,...,\alpha_l}(u_1,...,u_{l-1})\dif u_1,..,\dif u_{l-1}\dif t.
	\end{multline*}
Recognising that the product of Kronecker deltas is either zero or one gives
	\begin{multline}
	\label{eqap3}
	{\rm cum}\{w^{*\sT}_{i_{1}}(\ta,\tb),...,w^{*\sT}_{i_k}(\ta,\tb)\}
	\leq \sum_{l=1}^k  \sum_{\alpha_1,...,\alpha_l = 1}^p \\ \int\cdots\int \left[\ta^{-1/2}\psi^{\sT}\left\{\left(t+u_1\right)/\ta\right\}\right]^{|v_1|}\cdots \left[\ta^{-1/2}\psi^{\sT}\left\{(t+u_{l-1})/\ta\right\}\right]^{|v_{l-1}|}\left\{\ta^{-1/2}\psi^{\sT}\left(t/\ta\right)\right\}^{|v_{l}|}\\\cdot r_{\alpha_1,...,\alpha_l}(u_1,...,u_{l-1})\dif u_1,..,\dif u_{l-1}\dif t.
	\end{multline}
 Using the fact that $\sum_{j=1}^l |v_j| = k$,  H\"{o}lder's inequality gives
	\begin{multline}
	\label{eqap4a}
	\int \left|\left[\psi^{\sT}\left(t+u_1\right)\right]^{|v_1|}\cdots \left[\psi^{\sT}\left(t+u_{l-1}\right)\right]^{|v_{l-1}|}\left[\psi^{\sT}\left(t\right)\right]^{|v_{l}|}\right| \dif t \\
	\leq \left(\int \left|\psi^{\sT}\left(t\right)\right|^k \dif t \right)^{|v_l|/k}\prod_{\beta=1}^{l-1}  \left(\int \left|\psi^{\sT}\left(t+u_l\right)\right|^k \dif t \right)^{|v_\beta|/k}  = \int \left|\psi^{\sT}\left(t\right)\right|^k \dif t = A_k T^{1-k/2}
	\end{multline}
	where $A_k = \int|\psi(t)|^k\dif t$. Putting (\ref{eqap3}) and (\ref{eqap4a}) together, if follows that
	\begin{multline}
	\label{eqap4}
	|{\rm cum}\{w^{*\sT}_{i_{1}}(\ta,\tb),...,w^{*\sT}_{i_k}(\ta,\tb)\}| \\\leq (\ta T)^{1-k/2}A_k\sum_{l=1}^k \sum_{\alpha_1,...,\alpha_l=1}^p 
	\int\cdots\int \left|r_{\alpha_1,...,\alpha_l}(u_1,...,u_{l-1})\right|\dif u_1,..,\dif u_{l-1}.
	\end{multline}
	Therefore, from Assumption 3, ${\rm cum}\{w^{*\sT}_{i_{1}}(\ta,\tb),...,w^{*\sT}_{i_k}(\ta,\tb)\}$ is $O(T^{1-k/2})$ and tends to zero as $T\rightarrow\infty$ for all $k>2$. Therefore all cumulants of order greater than two asymptotically go to zero, giving the required asymptotic normality.
\end{proof}

\begin{theorem}
		Let $N(t)$ be a $p$-dimensional stationary process satisfying Assumption 3 with spectral density matrix $S(f)$. Let $\psi(t)$ be a wavelet with central frequency $f_0$ satisfying Assumption 4, let $h(t)$ be a smoothing function satisfying Assumption5, and for $\kappa>0$ let $\{\eta_l;\ l=0,1,\}$ be the eigenvalues of the kernel $K(s,t)$ defined in (4). The temporally smoothed wavelet periodogram $\Omega^{\sT}(\ta,\tb)$ is asymptotically $(1/n)\mathcal{W}^{\mathcal{C}}_p\{n,S(f_{\ta})\}$ as $T\rightarrow\infty$ for all $(\ta,\tb)\in \tilde{\mathcal{T}}_{\alpha,\kappa}$, where $n=1/\left(\sum_{l=1}^\infty\eta_l^2\right)$. 
\end{theorem}	
\begin{proof}
	Consider the eigenwavelet representation of the temporally smoothed wavelet periodogram in (5). By Lemma 3 and Theorem 1, $v_l^{\sT}(a,b)$ is asymptotically $\mathcal{N}\{0,S(f_0/\ta)\}$, $l=0,1,...$. Therefore, $v_l^{\sT}(\ta,\tb)v_l^{\sT}(\ta,\tb)^{\rm H}$ is asymptotically $\mathcal{W}_p^\mathcal{C}\{1,S(f_0/\ta)\}$. As the eigenwavelet system is orthonormal, Lemma \ref{lem:2} states that $v^{\sT}_0(\ta,\tb),v^{\sT}_1(\ta,\tb),...$ are asymptotically independent random vectors and therefore it follows in an analogous manner to \cite[p. 776]{Walden2000} that $\Omega^{\sT}(\ta,\tb)$ is asymptotically $(1/n)\mathcal{W}_p^{\mathcal{C}}\{n,S(f_0/\ta)\}$ where $n = 1/\sum_{l=0}^\infty\eta_l^2$.
\end{proof}

\begin{proposition}
	\label{prop4}
		Let $\psi(t)$ satisfying Assumption 4, let $h(t)$ be the rectangular smoothing window given in (6), and for $\kappa>0$ let corresponding kernel $K(s,t)$ have ordered eigenvalues $\{\eta_{l};\  l=0,1,...\}$. Provided $\kappa>\alpha$, then
	$
	n = (\sum_{l=0}^{\infty}\eta_l^2)^{-1} = \kappa \left\{\int^{\infty}_{-\infty} |\mathcal{P}(x)|^2\dif x\right\}^{-1} ,
	$ 
	where $\mathcal{P}(x) \equiv \int_{-\infty}^\infty\psi(t)\psi^*(t-x)\dif t$.
\end{proposition}
\begin{proof}
	It holds that 
\begin{equation}
	\int\int K(s,t)K^*(s,t)\dif s\dif t  = \int\int\left\{\sum_{l=0}^{\infty}\eta_l\varphi_l(s)\varphi^*_l(t)\right\}\left\{\sum_{l'=0}^{\infty}\eta_{l'}\varphi^*_{l'}(s)\varphi_{l'}(t)\right\}\dif s\dif t = \sum_{l=0}^{\infty}\eta_l^{2} \label{link}
\end{equation}
by the orthogonality of the eigenfunctions $\{\varphi_{l}(t);\  l=0,1,...\}$. Now,
\begin{align*}
\int\int K(s,t)K^*(s,t)\dif s\dif t & = \int\int\int\int h_\kappa(u) h_\kappa(v) \psi(s-u)\psi^*(t-u)\psi^*(s-v)\psi(t-v)\dif s\dif t\dif u\dif v \\
& = \frac{1}{\kappa^2}\int_{-\kappa/2}^{\kappa/2}\int_{-\kappa/2}^{\kappa/2}\int\int  \psi(s-u)\psi^*(s-v)\psi^*(t-u)\psi(t-v)\dif s\dif t\dif u\dif v.
\end{align*}
Considering individual integrals, we have
$$
 \int  \psi(s-u)\psi^*(s-v)\dif s = \int  \psi(s)\psi^*(s-(v-u))\dif s = \mathcal{P}(v-u)
$$
where $\mathcal{P}(x) \equiv \int\psi(t)\psi^*(t-x)\dif t$. Therefore 
\begin{align*}
\int\int K(s,t)K^*(s,t)\dif s\dif t & = 
 \frac{1}{\kappa^2}\int_{-\kappa/2}^{\kappa/2}\int_{-\kappa/2}^{\kappa/2}|\mathcal{P}(v-u)|^2\dif u\dif v \\
 & =  \frac{1}{2\kappa^2}\int_{-\kappa}^{\kappa}\int_{-\kappa}^{\kappa}|\mathcal{P}(x)|^2\dif x\dif y \\
 & = \frac{1}{\kappa}\int_{-\kappa}^{\kappa}|\mathcal{P}(x)|^2\dif x.
\end{align*}
The (approximating) support of $\psi(t)$ is $(-\alpha/2,\alpha/2)$, therefore the (approximating) support of $|\mathcal{P}(x)|^2$ is $(-\alpha,\alpha)$. Hence, provided $\alpha<\kappa$, it follows from (\ref{link}) that 
	$$
	n = \left(\sum_{l=0}^{\infty}\eta_l^2\right)^{-1} = \kappa \left\{\int |\mathcal{P}(x)|^2\dif x\right\}^{-1} .
	$$ 
\end{proof}

\subsection*{Test for non-stationarity: \\ Proofs of Propositions 5-7, and Theorem 3.}
\begin{proposition}
  Let $\psi(t)$ satisfy Assumption 4 and $h(t)$ satisfy Assumption 5. Then, for any $\kappa>0$ and $j>0$, $\Omega^{\sT}(\ta_j,\tb_{j,1}),...,\Omega^{\sT}(\ta_j,\tb_{j,2^j})$ are asymptotically independent.
\end{proposition}

\begin{proof}
	The kernels centred at $(\ta_j,\tb_{j,l})$ and $(\ta_j,\tb_{j,m})$ ($l\neq m$) are non-overlapping by construction of the dyadic partition. Hence, the eigen-wavelets at $(\ta_j,\tb_{j,l})$ are orthogonal to the eigenwavelets at $(\ta_j,\tb_{j,m})$. Lemma 2 combined with a similar argument to Theorem 1 gives the desired result.
\end{proof}

\begin{proposition}
	\label{lemma1}
	Let $B_1,...,B_K$ be independent samples where $B_i\sim(1/n)\mathcal{W}^{\mathcal{C}}_p(n,\Sigma_i)$ ($i=1,...,K$). The likelihood ratio test statistic for the null hypothesis $H:\Sigma_1=...=\Sigma_K=\Sigma$, with unspecified $\Sigma$, is 
$$
\tilde\Lambda = K^{pK n}\frac{\prod_{i=1}^{K}{\rm det}(B_i)^n}{{\rm det}\left(\sum_{i=1}^KB_i\right)^{Kn}}.
$$
Furthermore, when $H$ is true, $-2\log (	\tilde\Lambda)$ is asymptotically $\chi^2_{f}$ where $f  = (K-1)p^2$.
\end{proposition}

\begin{proof}
The probability density function for $B_i\sim \mathcal{W}^{\mathcal{C}}_p(n,\Sigma_i)$ is given as 
$$
f(B_i) = \frac{\det (B_i)^{n-p}\re^{-\tr (\Sigma^{-1}_i B_i)}}{\det(\Sigma_i)^n \Gamma^{\mathcal{C}}_p(n)},
$$
where $\Gamma^{\mathcal{C}}_p(\cdot)$ is the complex multivariate Gamma function. Apart from the multiplicative constant, the likelihood function based on the independent samples $B_1,...,B_K$ is  
$$
L(\Sigma_1,...,\Sigma_K)  = \prod_{i=1}^K \frac{\det (B_i)^{n-p}\re^{-\tr (\Sigma^{-1}_i B_i)}}{\det(\Sigma_i)^n}, 
$$
and the  likelihood ratio test statistic is defined as 
\begin{equation}
\label{lambdadef}
\tilde\Lambda \equiv \frac{\sup_{\Sigma>0}L(\Sigma,...,\Sigma)}{\sup_{\Sigma_1,...,\Sigma_K>0}L(\Sigma_1,...,\Sigma_K)}  = \frac{L(n^{-1}\bar B,...,n^{-1}\bar B)}{L(n^{-1}B_1,...,n^{-1}B_K)}, 
\end{equation}
where $\bar B = (1/K)\sum_{i=1}^{K}B_i$. Hence the numerator in (\ref{lambdadef}) is 
\begin{align*}
L(n^{-1}\bar B,...,n^{-1}\bar B) & = \prod_{i=1}^K \frac{\det (n^{-1}B_i)^{n-p}\re^{-\tr (n\bar B^{-1} B_i)}}{\det(n^{-1}\bar B)^n} \\
& = \frac{(nK)^{pnK}\re^{-pnK}}{\det\left(\sum_{i=1}^K B_i\right)^{nK}}\prod_{i=1}^K \det(n^{-1}B_i)^{n-p},
\end{align*}
and the denominator is 
\begin{align*}
L(n^{-1}B_1,...,n^{-1}B_K) & = \prod_{i=1}^K \frac{\det (n^{-1}B_i)^{n-p}\re^{-\tr (nB_i^{-1} B_i)}}{\det(n^{-1}B_i)^n} \\
& = n^{pnK}\re^{-pnK}\prod_{i=1}^K \frac{\det(n^{-1}B_i)^{n-p}}{\det\left(\sum_{i=1}^K B_i\right)^{n}}.
\end{align*}


Therefore 
$$
\tilde\Lambda = K^{pK n}\frac{\prod_{i=1}^{K}{\rm det}(B_i)^n}{{\rm det}\left(\sum_{i=1}^KB_i\right)^{Kn}}.
$$
The asymptotic distribution follows from the general theory of likelihood ratio tests.
\end{proof}

\begin{proposition}
	Let $\tk = \kappa T^{c}$ where $\kappa>0$ and $0<c<1/2$, and define the test statistic for $H_j$ as
$$
\Lambda_j = K^{pK n}\frac{\prod_{i=1}^{K}{\rm det}\{\Omega^{\sT}(\ta_j,\tb_i)\}^n}{{\rm det}\left\{\sum_{i=1}^K\Omega^{\sT}(\ta_j,\tb_i)\right\}^{Kn}},
$$
where $K = 2^j$ and $n$ is as given in Theorem 2. Under $H_j$, $\Lambda_j\equaldist 	\tilde\Lambda_0(\Sigma) + o(1)$, where $\Sigma = E\{\Omega(a,b)\}$.
\end{proposition}

\begin{proof}
	We consider the eigen-wavelet decomposition 
	$$
	\Omega^{\sT}(\ta,\tb) = \sum_{k=0}^{\infty}\eta_k v_k^{\sT}(\ta,\tb)v_k^{\sT}(\ta,\tb)^{\rm H},
	$$
	with $\sum_{k=0}^{\infty}\eta_k = 1$.
	It follows from Theorem 1 and the Cornish-Fisher inversion \cite[p. 117]{OEBN1989} that $\nu_k^{\sT}(\ta,\tb) \equaldist Z_k + R^{\sT}_k$, where $Z_k$ are iid $N_p^\mathcal{C}(0,\Sigma)$ and $R^{\sT}_k = O(T^{-1/2})$. Therefore,
	\begin{eqnarray*}
	 \Omega^{\sT}(\ta,\tb) &\equaldist& \sum_{k=0}^{\infty}\eta_k Z_kZ_k^{\rm H} + \sum_{k=0}^{\infty}\eta_k Z_kR^{{\sT} {\rm H}}_k + \sum_{k=0}^{\infty}\eta_k R^{\sT}_kZ_k^{\rm H} +  \sum_{k=0}^{\infty}\eta_k R^{\sT}_k R^{{\sT} {\rm H}}_k  \\
	&\equaldist& B + O(T^{-1/2}) + O(T^{-1/2}) + O(T^{-1}).  \\
	&\equaldist& B + O(T^{-1/2}).
	\end{eqnarray*}
	where $B\sim (1/n)\mathcal{W}^{\mathcal{C}}_p(n,\Sigma)$. Let $\Omega^{\sT}_{j,i}$ be shorthand for $\Omega^{\sT}(\ta_j,\tb_i)$. It is true that $\det(\Omega^{\sT}_{j,i}) = \det(B) + O(T^{-1/2})$. It follows that, under $H_0$
	\begin{align*}
	\log(\Lambda_j) & = pK n \log(K) + n\sum_{i=1}^K \log\left\{\det\left(\Omega^{\sT}_{j,i}\right)\right\} - Kn\log\left\{\det\left(\sum_{i=1}^K\Omega^{\sT}_{j,i}\right)\right\} \\
	& \equaldist pK n \log(K) + n\sum_{i=1}^K \log\left\{\det(B) + O(T^{-1/2})\right\} + Kn\log\left[K\{\det(B) + O(T^{-1/2})\}\right] \\
	& \equaldist pK n \log(K) + n\sum_{i=1}^K \left[\log\{\det(B)\} + O(T^{-1/2})\right] + Kn\log\{K\det(B)\} + O(T^{-1/2}).
	\end{align*}
From Proposition \ref{prop4}, if $\kappa = O(T^c)$, then so too is $n=O(T^c)$. Therefore 
\begin{align}
\log(\Lambda_j) & \equaldist pK n \log(K) + n\sum_{i=1}^K \log\{\det(B)\} + Kn\log\{K\det(B)\}+ O(T^{c-1/2})\nonumber \\
& \equaldist \log\{\tilde\Lambda_0(\Sigma)\} + O(T^{c-1/2}).\nonumber
\end{align}
Therefore \begin{equation}\label{loglambda}\Lambda_j \equaldist \tilde\Lambda_0(\Sigma)\{1 + O(T^{c-1/2})\},\end{equation} and hence $\Lambda_j  = \tilde\Lambda_0(\Sigma) + o(1)$ if $0<c<1/2$.
\end{proof}

\begin{theorem}
	Let $\tk = \kappa T^{c}$ where $\kappa>0$ and $0<c<1$. Under $H_j$, $-2\log (\Lambda_j)$ is asymptotically $\chi^2_{\nu_j}$ where $\nu_j  = (2^j-1)p^2$. Specifically, ${\rm pr}\{-2\log (\Lambda_j)\leq x\} = {\rm pr}(\chi^2_{\nu_j}\leq x) + O(T^{-\beta})$ where $\beta = \min\{c,1/2-c\}$.
\end{theorem}

\begin{proof}
	It is true that $E(\Lambda_j^k) = E[\tilde\Lambda_0(\Sigma)^k\{1+O(T^{c-1/2})\}^k] = E[\tilde\Lambda_0(\Sigma)^k\{1+O(T^{c-1/2})\}]$. Let $Y  = -2\log(\Lambda_j)$, then the characteristic function of $Y$ is given as
	\begin{align}
	\phi_{Y}(t) = E(\re^{\ri t Y}) = E(\re^{-2\ri t \log\Lambda_j}).
	\end{align}
	From Equation (\ref{loglambda}), we have 
	$$
	\phi_Y(t) = E\{\tilde\Lambda_0(\Sigma)^{ -2\ri t}\}\{1+O(T^{c-1/2})\} = \phi_{\tilde Y_0}(t)\{1+O(T^{c-1/2})\},
	$$
	where $\tilde Y_0  = -2\log\{\tilde\Lambda_0(\Sigma)\}$. From \cite[p. 306]{muirhead1985}, $\phi_{\tilde Y_0}(t) = (1-2\ri t)^{-\nu_j/2}\{1+O(T^{-c})\}$, and hence
	\begin{align*}
	\phi_{Y}(t) & = (1-2\ri t)^{-\nu_j/2}\{1+O(T^{-c})\}\{1+O(T^{c-1/2})\} \\
	& = (1-2\ri t)^{-\nu_j/2}\{1+O(T^{-c}) + O(T^{c-1/2})  + O(T^{-1/2})\}.
	\end{align*}
	Therefore, for $0<c<1/2$, it follows that $\phi_{Y}(t) = (1-2\ri t)^{-\nu_j/2}\{1+O(T^{-\beta})\}$ where $\beta = \min\{c,1/2-c\}$. It directly follows that ${\rm pr}\{-2\log (\Lambda_j)\leq x\} = {\rm pr}(\chi^2_{\nu_j}\leq x) + O(T^{-\beta})$.

\end{proof}

\section{Real valued wavelets}

The results for real valued wavelets are extremely similar to the complex valued wavelet setting. Specifically, a similar line of argument results in $w^{\sT}(\ta,\tb)$ being asymptotically $\mathcal{N}_p\{0,S(f_{\ta})\}$ as $T\rightarrow\infty$, for all $(\ta,\tb)\in\tilde{\mathcal{T}}_{\alpha,\kappa}$. The condition placed on complex-valued wavelets to be orthogonal to their complex conjugate clearly does not hold for real valued wavelets, however it is only needed in the complex setting to obtain a circular complex-Gaussian distribution, which is not needed for the real valued wavelet.

\begin{figure}[t!]
	\begin{center}\includegraphics[width=0.7\linewidth]{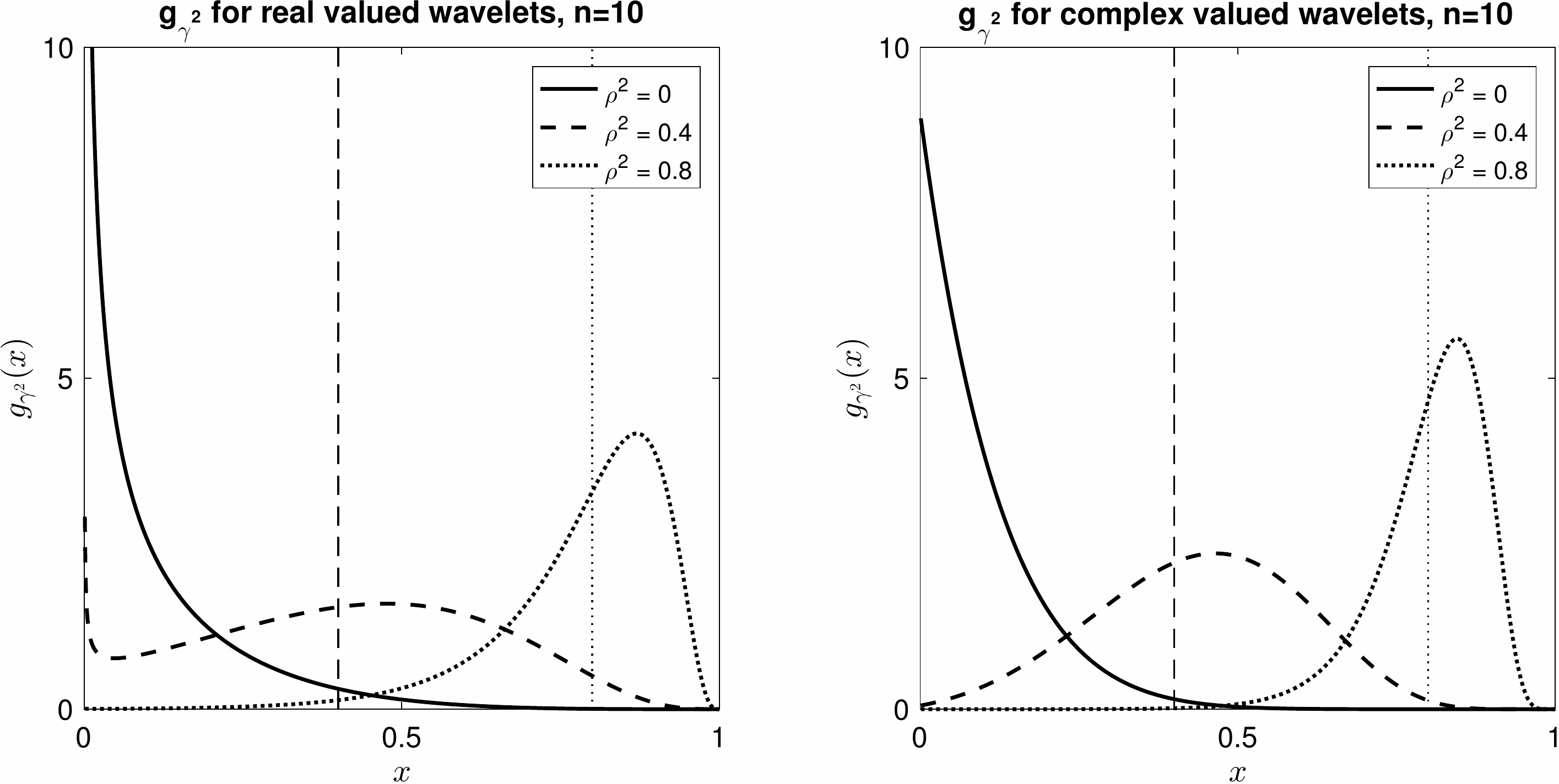}\caption{\label{pdfs}Asymptotic probability density functions for temporally smoothed wavelet coherence using real and complex valued wavelets. Degrees of freedom $n=10$ and true coherence $\rho^2=0$, $0.4$, and $0.8$ (marked with the verticle lines)}
	\end{center}
\end{figure}

It then follows that $\Omega^{\sT}(\ta,\tb)$ is asymptotically $(1/n)\mathcal{W}_p\{n,S(f_{\ta})\}$ as $T\rightarrow\infty$ for all $(\ta,\tb)\in \tilde{\mathcal{T}}_{\alpha,\kappa}$, where $n=1/\left(\sum_{l=1}^\infty\eta_l^2\right)$, again along a similar line of argument. The distribution to the wavelet coherence is slightly different for real valued wavelets. Let $\Omega^{\sT ij}(a,b)$ be the $\mathbb{R}^{2\times 2}$ matrix made up of the $i$th and $j$th columns and rows of $\Omega^{\sT}(a,b)$. It is immediate that asymptotically $\Omega^{\sT ij}\sim \mathcal{W}_2\{n,S(f_{\tilde a})\}$ and from Theorem 5.3.2 of \cite{muirhead1985} (first presented in \cite{Fisher1928}) that $\gamma^2_{ij}(\tilde a,\tilde b)$ asymptotically has density function
$$
g_{\gamma^2}(x) =	\frac{\Gamma(\frac{1}{2}n)}{\Gamma(\frac{1}{2})\Gamma\{\frac{1}{2}(n-1)\}}	x^{-1/2}(1-x)^{(n-3)/2}(1-\rho^2)^{n/2}
\ _2 F_1(n/2,n/2,1/2,\rho^2x)
$$
where $\rho^2$ is shorthand for $\rho_{ij}^2(f_{\tilde a})$, the spectral coherence between $N_i(t)$ and $N_j(t)$ at frequency $f_{\tilde a}$. In the case of $\rho_{ij}^2(f_{\tilde a}) = 0$, this distribution is Beta$\{1/2,(n-1)/2\}$. The density functions are shown in Figure \ref{pdfs}.

The test of stationarity at scale $j$ is also extensible to real valued wavelets. However, the test statistic becomes
	$$
\Lambda_j = K^{pK n/2}\frac{\prod_{i=1}^{K}{\rm det}\{\Omega^{\sT}(\ta_j,\tb_i)\}^{n/2}}{{\rm det}\left\{\sum_{i=1}^K\Omega^{\sT}(\ta_j,\tb_i)\right\}^{Kn/2}}.
$$
The asymptotic distribution of $\Lambda_j$ under $H_j$ and the rate of convergence is identical to complex valued wavelets.

\section{Simulation details}
To validate the distributional result for the wavelet transform in Theorem 1 (and the corresponding real valued wavelet result), two univariate processes were used. The first is a homogeneous Poisson process for which $\Gamma(\tau) = \lambda\delta(\tau)$ and $S(f) = \lambda$. The second is a Hawkes process \citep{Hawkes1971} with exponential decay. This self-exciting process contains internal correlation structure and has a stochastic intensity function of 
$$
\Lambda(t) = \nu + \int_{-\infty}^{t}\alpha \exp\{-\beta(t-s)\}\dif N(s),
$$
and a spectral density function
$$
S(f) = \frac{\nu\beta}{\beta-\alpha}\left\{1+\frac{ \alpha(2\beta-\alpha)}{(\beta-\alpha)^2 + (2\pi f)^2}\right\}.
$$

\begin{figure}[h!]
	\begin{center}\includegraphics[width=0.8\linewidth]{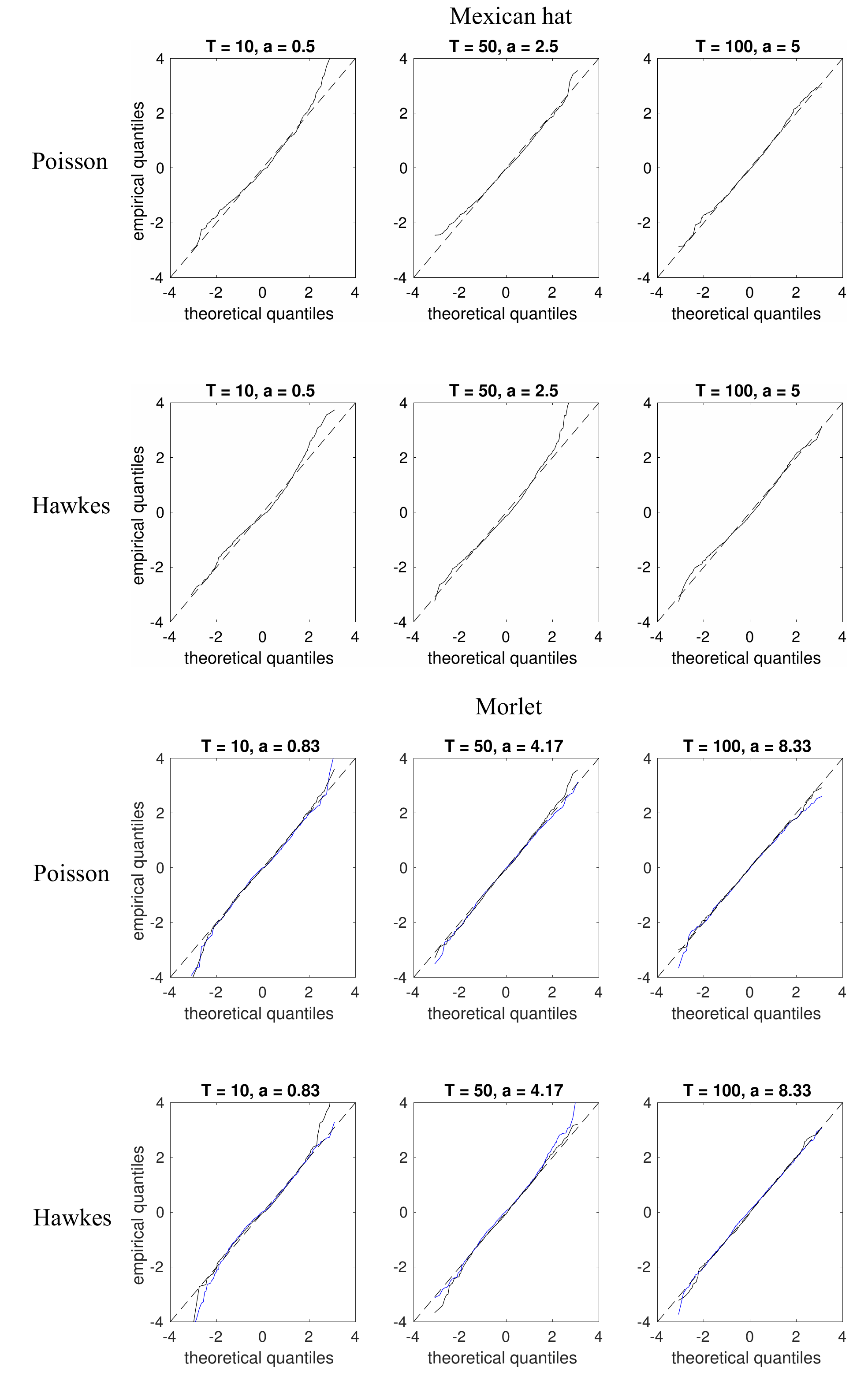}\caption{\label{qqplots}QQ plots for empirical quantiles of the wavelet transform against the theoretical quantiles of the normal distribution. See text for details.}
	\end{center}
\end{figure}

The Hawkes process used in the simulations has a base-intensity $\nu = 1$, excitation intensity $\alpha = 0.5$, and decay parameter $\beta=1$. Both the real valued Mexican hat wavelet and the complex valued Morlet wavelet are considered for $T=10$, 50 and 100. The asymptotic normality of the wavelet transform, as presented in Theorems 1 and 3, are confirmed via QQ plots in Figure \ref{qqplots}. Here, the ordered values of $\sigma^{-1}w(a,b)$ ($\sigma^2 = S(f_0/a)$) are plotted against the theoretical quantiles of the standard normal.

To validate the distributional results of the temporally smoothed wavelet coherence in Theorem 2 and Supplementary Material Section 2, two bivariate processes are used. The first is a pair of independent homogeneous Poisson processes. Its spectral density matrix is 
$$
S(f) = \left(\begin{array}{cc}
\lambda & 0 \\ 
0 & \lambda
\end{array} \right)
$$
and has a true spectral coherence of $\rho^2(f) = 0$.

The second is a bivariate mutually exciting Hawkes process. This contains both inter and cross dependencies and is defined through its stochastic intensity function
$$
\Lambda(t) = \nu + \int_{-\infty}^{t}G(t-s)\dif N(s),
$$
where $\nu$ is the base intensity vector and 
$$
G(s) = \left(\begin{array}{cc}
\alpha_{11}\exp(-\beta_{11}s) & \alpha_{12}\exp(-\beta_{12}s) \\ 
\alpha_{12}\exp(-\beta_{21}s) & \alpha_{22}\exp(-\beta_{22}s)
\end{array} \right).
$$
The diagonal elements of $G(s)$ characterise the self-exciting behaviour of each individual process, and the off-diagonal elements characterise the mutually exciting behaviour. The particular process used in the simulations has $\nu=(1,1)^T$, $\beta_{ij} = 1$ ($i,j=1,2$), $\alpha_{11} = \alpha_{22} = 0.5$, and $\alpha_{12} = \alpha_{21} = 0.4$.

\begin{figure}
	\begin{center}\includegraphics[width=0.4\linewidth]{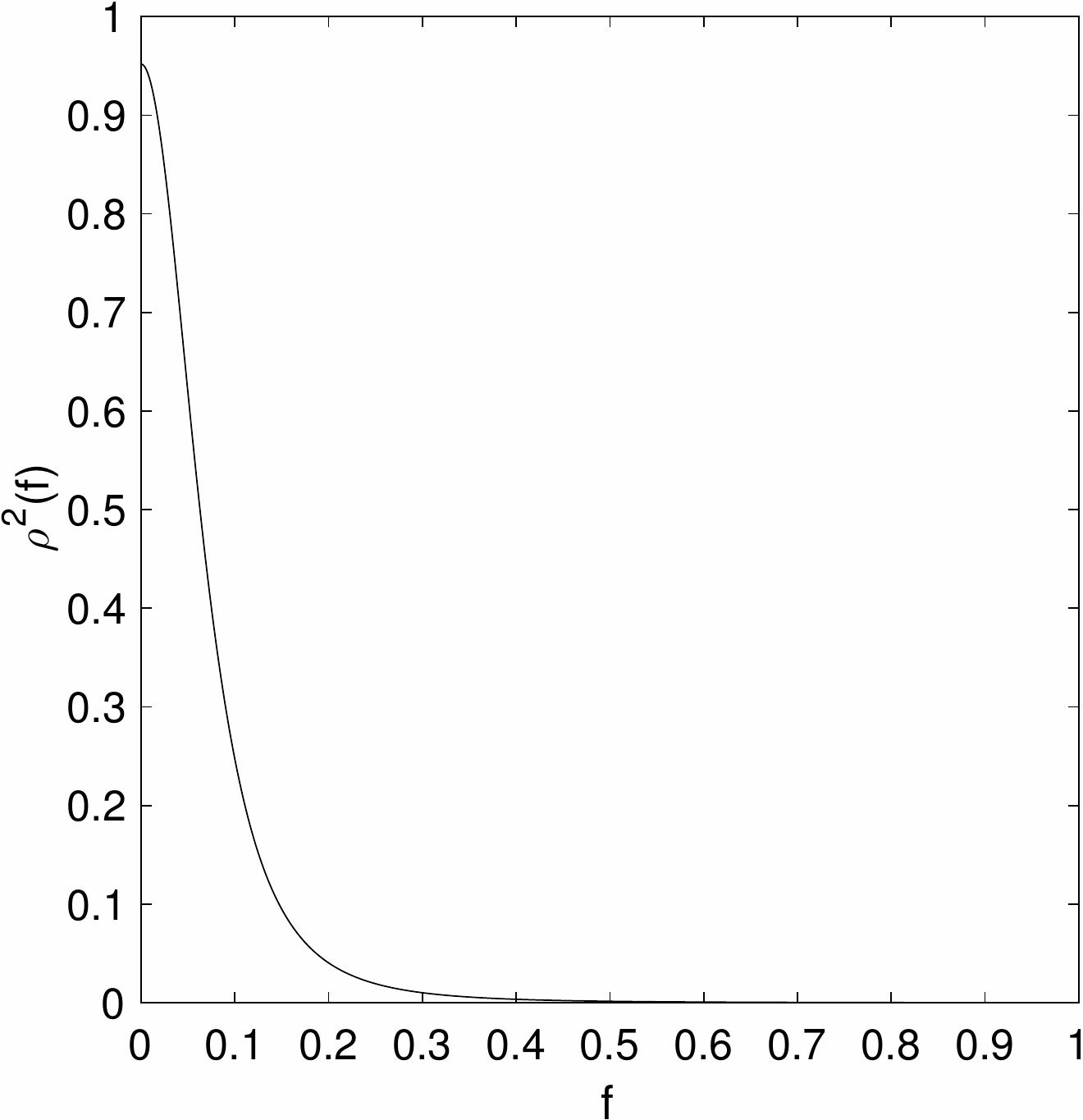}\caption{\label{speccoh}Spectral coherence $\rho^2(f)$ for the bivariate mutually exciting Hawkes process described in the text.}
	\end{center}
\end{figure}

The form of the spectral matrix in this setting is non-trivial and given in \cite{Hawkes1971}. The true spectral coherence can be computed from this matrix and is shown in Figure \ref{speccoh}.

\begin{figure}
	\begin{center}\includegraphics[width=0.8\linewidth]{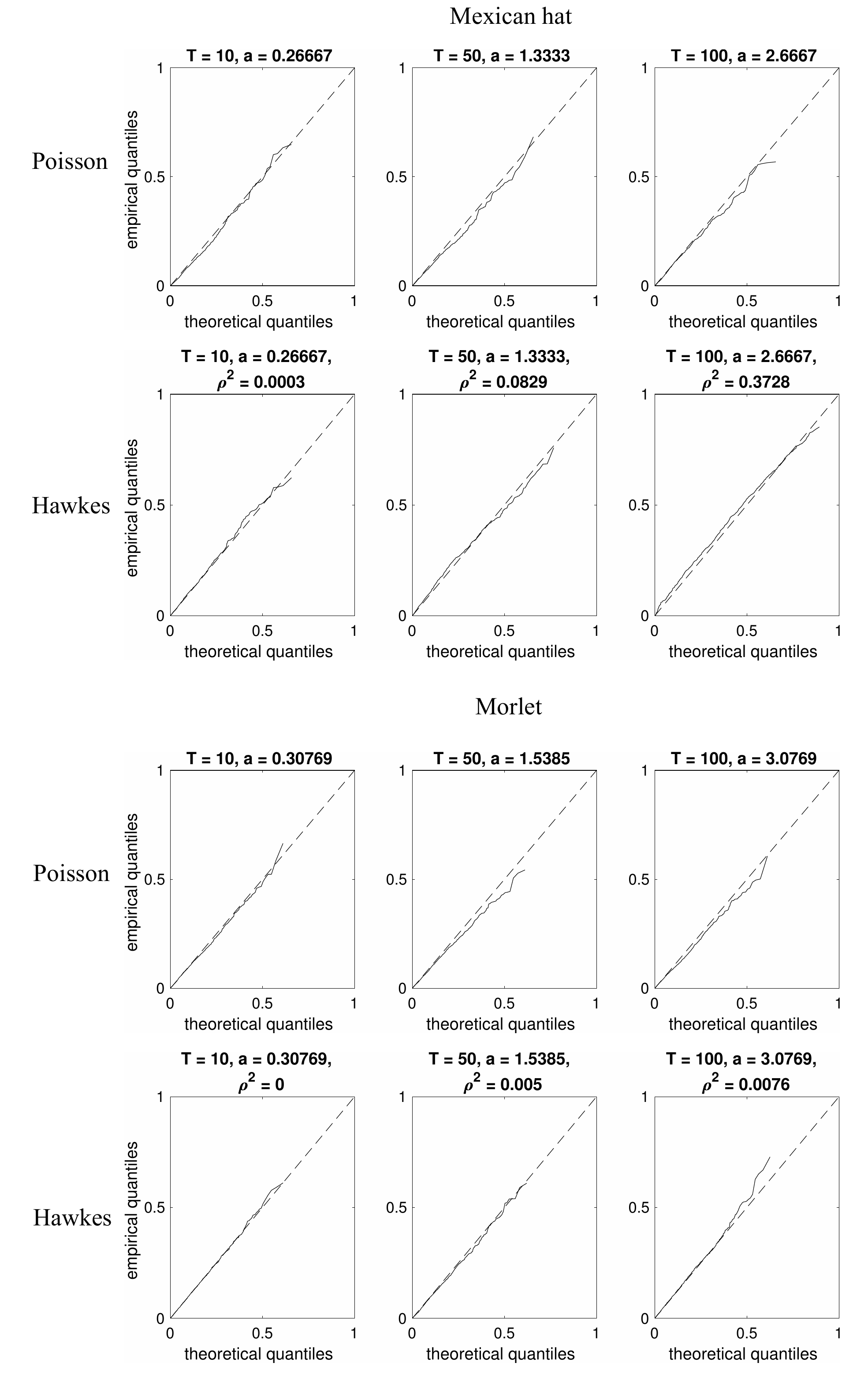}\caption{\label{cohqqplots}QQ plots for empirical quantiles of the temporally smoothed wavelet coherences against the theoretical quantiles of the asymptotic distribution. See text for details.}
	\end{center}
\end{figure}

The asymptotic distribution for the temporally smoothed wavelet coherence are confirmed via QQ plots in Figure \ref{cohqqplots} for both the Poisson and Hawkes processes. Here, the ordered values of $\gamma_{12}^2(a,b)$ are plotted against the theoretical quantiles of the stated distributions. Both the real valued Mexican hat wavelet and the complex valued Morlet wavelet are considered at scale $\ta = 4/5$ for $T=10$, 50 and 100. Translation $\tb=1/2$ and a smoothing parameter of $\kappa = 20$ is used, resulting in 11.57 degrees of freedom for the Mexican hat wavelet and 8.31 degrees of freedom for the Morlet wavelet. 

\begin{figure}[t!]
	\begin{center}\includegraphics[width=1\linewidth]{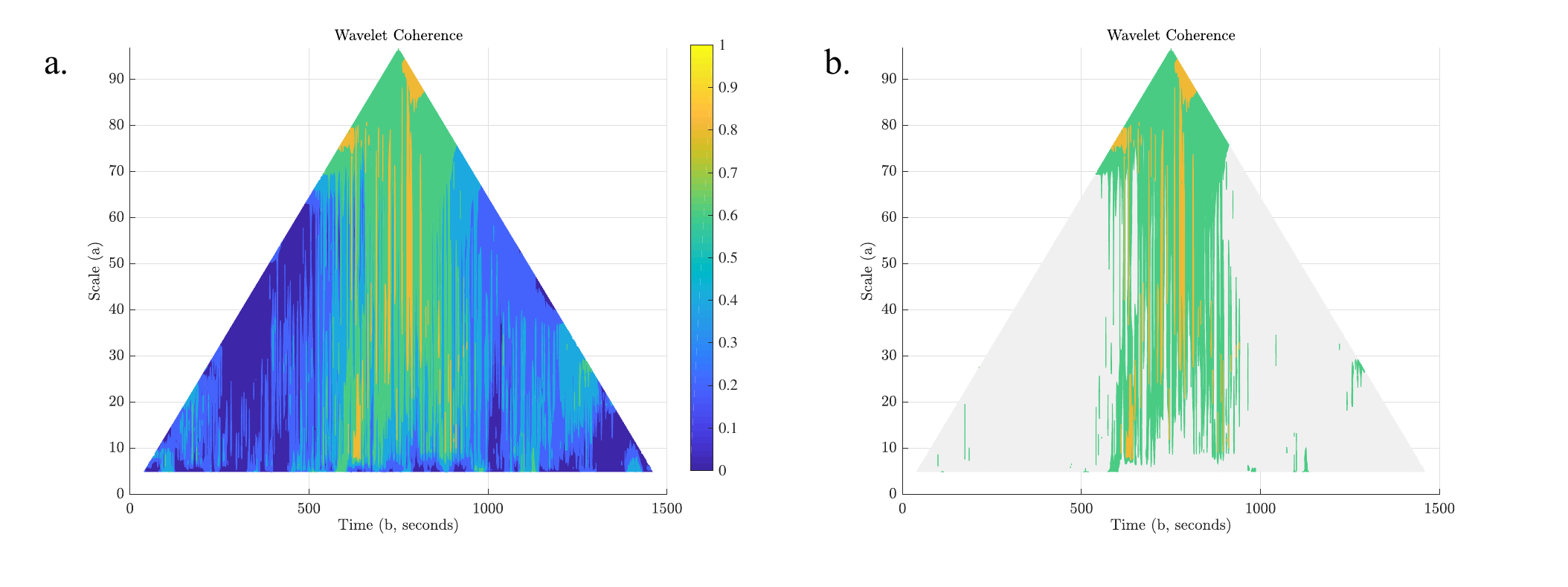}\caption{\label{hawkes} a. Temporally smoothed wavelet coherence estimated from a single realisation of a piecewise stationary bivariate Hawkes process. b. Temporally smoothed wavelet coherence values above the 95th percentile of the null (zero coherence) distribution.}
	\end{center}
\end{figure}

To demonstrate the ability of temporally smoothed wavelet coherence to detect non-stationary coherence, we compose a piecewise stationary bivariate point process. On the intervals $(0,500]$ and $(1000,1500]$ are a pair independent Hawkes processes each with parameters $\nu = (0.5,0.5)^T$, $\alpha = 0.7$ and $\beta = 1$. On the interval $(500,1000]$ the processes switch to being mutually exciting Hawkes processes with positive coherence. Specifically we set $\nu=(0.5,0.5)^T$, $\beta_{ij} = 1$ ($i,j=1,2$), $\alpha_{11} = \alpha_{22} = 0.2$, and $\alpha_{12} = \alpha_{21} = 0.5$. Fig. \ref{hawkes}a shows the temporally smoothed wavelet coherence for a single realisation of the described process using a Morlet wavelet and smoothing parameter $\kappa = 10$. Fig. \ref{hawkes}b highlights those regions of the time-scale space for which the coherence is larger than the 95th percentile of the null distribution (0.593), clearly demonstrating its ability to detect non-stationary coherence.
The bivariate piecewise stationary Hawkes process used in Section 4.4 is made up of three components. On the intervals $(0,500]$ and $(1000,1500]$ are a pair independent Hawkes processes each with parameters $\nu = (0.5,0.5)^T$, $\alpha = 0.7$ and $\beta = 1$. On $(500,1000]$ is a bivariate mutually exciting Hawkes process with $\nu=(0.5,0.5)^T$, $\beta_{ij} = 1$ ($i,j=1,2$), $\alpha_{11} = \alpha_{22} = 0.2$, and $\alpha_{12} = \alpha_{21} = 0.5$.

\section{Code}
Code and data associated with this paper can be found in the github repository 

https://github.com/AlexGibberd/pointWav.

\bibliographystyle{chicago}
\bibliography{bib}